\documentclass[floatfix,lengthcheck,showpacs,amssymb,amsmath,amsfonts,twocolumn]{aastex63}
\usepackage{CJK}
\usepackage[utf8]{inputenc}
\usepackage{color}
\usepackage{graphicx}
\usepackage{float}
\usepackage{subfigure}
\usepackage{amsmath}
\usepackage{afterpage}
\usepackage{acro}
\usepackage{multirow}
\usepackage{tikz}
\usetikzlibrary{arrows,shapes,trees,decorations.pathreplacing}
\usepackage{import}
\usepackage{svn-multi}
\usepackage{hyperref}
\usepackage{gensymb}
\usepackage{longtable}

\hypersetup{
    colorlinks=true,
    linkcolor=blue,
    filecolor=magenta,      
    urlcolor=cyan,
}

\DeclareAcronym{PSD}{
	short = PSD,
	long  = power spectral density
}

\newcommand{\msun}{\ensuremath{\mathrm{M}_{\odot}}}

\usepackage{xcolor,colortbl}
\definecolor{o1}{gray}{0.9}
\definecolor{o2}{gray}{0.8}
\definecolor{o3}{gray}{0.6}
\definecolor{Gray}{gray}{0.9}
\usepackage{longtable}

\begin{document}
\begin{CJK*}{UTF8}{gbsn}
\title[]{4-OGC: Catalog of gravitational waves from compact-binary mergers}

\correspondingauthor{Alexander H. Nitz}
\email{alex.nitz@aei.mpg.de}

\author[0000-0002-1850-4587]{Alexander H. Nitz}
\author[0000-0002-6404-0517]{Sumit Kumar}
\author[0000-0002-2928-2916]{Yi-Fan Wang (王一帆)}
\author[0000-0003-0966-1748]{Shilpa Kastha}
\author[0000-0002-9188-5435]{Shichao Wu (吴仕超)}
\author[0000-0002-6990-0627]{Marlin Sch{\"a}fer}
\author[0000-0002-5077-8916]{Rahul Dhurkunde}
\affil{Max-Planck-Institut f{\"u}r Gravitationsphysik (Albert-Einstein-Institut), D-30167 Hannover, Germany}
\affil{Leibniz Universit{\"a}t Hannover, D-30167 Hannover, Germany}
\author[0000-0002-0355-5998]{Collin D. Capano}
\affil{Max-Planck-Institut f{\"u}r Gravitationsphysik (Albert-Einstein-Institut), D-30167 Hannover, Germany}
\affil{Leibniz Universit{\"a}t Hannover, D-30167 Hannover, Germany}
\affil{Department of Mathematics, University of Massachusetts, Dartmouth, MA 02747, USA}


\begin{abstract}
We present the fourth Open Gravitational-wave Catalog (4-OGC) of binary neutron star (BNS), binary black hole (BBH) and neutron star-black hole (NSBH) mergers. The catalog includes observations from 2015-2020 covering the first through third observing runs (O1, O2, O3a, O3b) of Advanced LIGO and Advanced Virgo. The updated catalog includes 7 BBH mergers which were not previously reported with high significance during O3b for a total of 94 observations: 90 BBHs, 2 NSBHs, and 2 BNSs.
The most confident new detection, GW200318\_191337, has component masses $49.1^{+16.4}_{-12.0}\msun$ and $31.6^{+12.0}_{-11.6}\msun$; its redshift of $0.84^{+0.4}_{-0.35}$ ($90\%$ credible interval) may make it the most distant merger so far. We estimate the merger rate of BBH sources, assuming a powerlaw mass distribution containing an additive Gaussian peak, to be $16.5_{-6.2}^{+10.4} (25.0_{-8.0}^{+12.6})$ Gpc$^{-3}$ yr$^{-1}$ at a redshift of $z=0$ (0.2). For BNS and NSBH sources, we estimate a merger rate of  $200^{+309}_{-148}$ Gpc$^{-3}$ yr$^{-1}$ and $19^{+30}_{-14}$ Gpc$^{-3}$ yr$^{-1}$, respectively, assuming the known sources are representative of the total population.
We provide reference parameter estimates for each of these sources using an up-to-date model accounting for instrumental calibration uncertainty. The corresponding data release also includes our full set of sub-threshold candidates. 
\end{abstract}

\section{Introduction}
Gravitational-wave astronomy has entered an era of regular and routine observation of compact-binary mergers. This achievement was made possible by the second-generation gravitational-wave observatories, led by the twin Advanced LIGO (Hanford and Livingston)~\citep{TheLIGOScientific:2014jea} and Advanced Virgo~\citep{TheVirgo:2014hva} observatories, which have been operating since 2015 and 2017, respectively. This period has seen continued improvement in their astrophysical reach over their three completed observing runs (O1-O3)~\citep{Abbott:2020qfu}, with the pace rapidly increasing from 3 merger observations in O1 to dozens in the first half of O3 (O3a)~\citep{LIGOScientific:2018mvr,Abbott:2020niy,LIGOScientific:2021usb, Venumadhav:2019lyq, Nitz:2019hdf,Nitz:2021uxj}; the vast majority of these are binary black hole mergers (BBHs). To date, there is only a single binary neutron star (BNS) observation, GW170817~\citep{TheLIGOScientific:2017qsa}, which has been corroborated by extensive electromagnetic observations~\citep{GBM:2017lvd}. In addition, GW190425 is a potential heavy BNS merger~\citep{Abbott:2020uma} and recently two sources with masses compatible with merging neutron star-black hole binaries have been reported~\citep{LIGOScientific:2021qlt}. The plethora of BBH observations, in addition to exceptional events such as GW190521 with total mass $\sim 150 \msun$ ~\citep{Abbott:2020tfl,Capano:2021etf}, are beginning to constrain formation scenarios~\citep{Abbott:2020mjq,Gerosa:2021mno,Edelman:2021fik,Zevin:2020gbd} and deviations from general relativity~\citep{LIGOScientific:2020tif,Wang:2021gqm}.

We expect the current observatories to continue to improve in sensitivity over the next few years~\citep{Abbott:2020qfu} due to ongoing active commissioning and the inclusion of new technologies into the existing sites~\citep{LIGOScientific:2013pcc}. The upcoming O4 observing run is scheduled to begin at the end of 2022 with a fiducial BNS range of 160-190 Mpc~\citep{Abbott:2020qfu}. In addition, we can expect that a fourth ground based gravitational-wave observatory, KAGRA, will join the O4 observing run~\citep{KAGRA2020}. In the mid-to-late 2020's a fifth observatory will join the worldwide network with the construction of LIGO India~\citep{Unnikrishnan:2013qwa,Saleem:2021iwi}. Efforts are also advancing for third-generation observatories, Cosmic Explorer and the Einstein Telescope, which promise an order of magnitude sensitivity improvement~\citep{cewhitepaper,Evans:2021gyd,et}.

This work provides a comprehensive catalog of gravitational-wave observations from merging BNS, BBH, and NSBH sources. The analysis is based on a deep archival search using the public data from the LIGO and Virgo observatories~\citep{Vallisneri:2014vxa,Abbott:2019ebz} which spans 2015-2020 and includes all existing observing runs (O1-O3). This catalog updates the results of our previous 3-OGC catalog~\citep{Nitz:2021uxj} by including analysis of data from the second half of the third observing run (O3b) which very recently became public. The next expected public data release would be at the end of 2024 if the current release delay of 18 months with a 6-month cadence is maintained. Included in our companion data release is the complete set of sub-threshold candidates and parameter estimates for significant candidates~\citep{4-OGC}. We make available our sub-threshold candidates so that they may aid follow-up studies, including those which cross-correlate candidates with other archival datasets such as from gamma-ray bursts~\citep{Burns:2018pcl,Nitz:2019bxt}, high-energy neutrinos~\citep{Countryman:2019pqq}, or optical transients~\citep{Andreoni:2018fcm,Setzer:2018ppg}. Archival analyses have the potential to uncover distant or faint populations.

We find a total of 94 mergers which pass our significance threshold, $\mathcal{P}_{\textrm{astro}} > 0.5$ or false alarm rate (FAR) less than once per 100 years. The vast majority, 90, of these are BBH mergers; 7 are reported during O3b for the first time here with high significance, 3 additional were previously reported marginal candidates in the 3-OGC catalog~\citep{Nitz:2021uxj}. We find the previously-reported BNS and NSBH mergers~\citep{TheLIGOScientific:2017qsa,Abbott:2020uma,LIGOScientific:2021qlt}, however, no new confident BNS or NSBH mergers are observed. Our results are broadly consistent with the recent GWTC-3 catalog produced by the LVK (LIGO-Virgo-KAGRA) collaborations~\citep{LIGOScientific:2021djp,LIGOScientific:2021psn}. 

\begin{figure*}[t]
  \centering
    \hspace*{-1.8cm}\includegraphics[width=2.6\columnwidth]{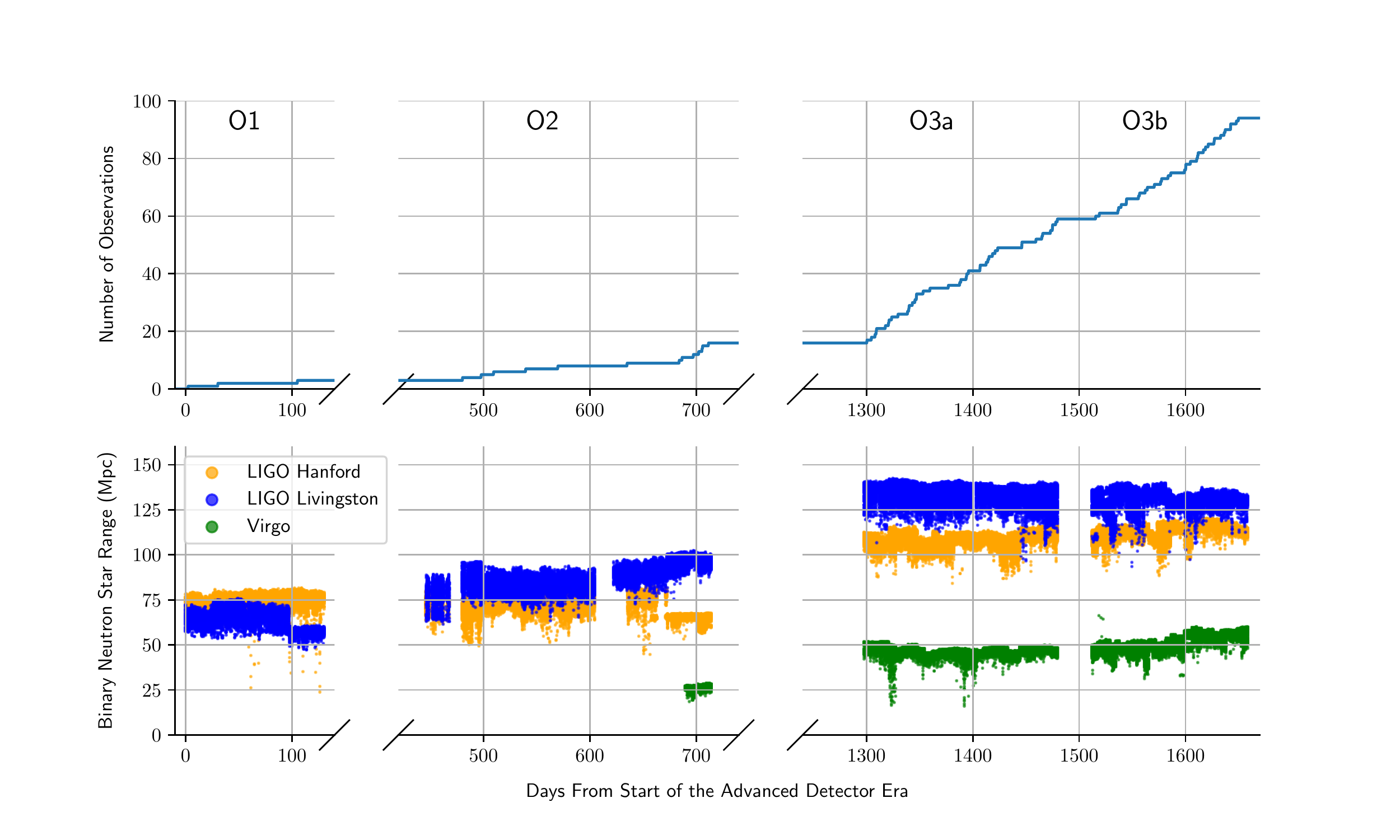}
\caption{The cumulative number of binary merger observations (top) and the fiducial 1.4-1.4 \msun BNS merger range (bottom) of the LIGO Hanford (yellow), LIGO Livingston (blue), and Virgo (green) observatories at a SNR of 8 as a function of days since the start of the advanced detector era. The distance is averaged over sky location and orientation angles. The O1 (left), O2 (middle) and O3 (right) observing periods are shown. The break in O3 demarks the boundary between O3a and O3b.}
\label{fig:range}
\end{figure*}

\begin{table}[b]
  \begin{center}
    \caption{Analyzed time in days for different global network observing scenarios. The abbreviations H, L, and V are used for the LIGO-Hanford, LIGO-Livingston, and Virgo observatories, respectively. Each time period is exclusive of the others. Some data is excluded from the full public dataset due to analysis requirements ($O(1)\%$). However, data around GW170608 and GW190814 which was made available separately from the bulk data release is included~\citep{Vallisneri:2014vxa}.}
    \label{table:data}
\begin{tabular}{rrrrrrrr}
Observation & HLV & HL & HV & LV & H & L & V \\\hline
O1 \vline& - & 48.6 & - & - & 27.6 & 17.0 & - \\
O2 \vline&15.2 & 103.3 & 1.7 & 2.2 & 37.8 & 33.0 & 1.7 \\
O3 \vline&152.0 & 49.5 & 31.7 & 38.9 & 10.3 & 9.9 & 25.0 \\
All \vline&167.2 & 201.4 & 33.4 & 41.1 & 75.6 & 59.9 & 26.7 \\
\end{tabular}
  \end{center}
\end{table}

\section{Search for Compact-binary Mergers}

Our catalog includes results from the analysis of the complete set of public LIGO and Virgo data collected over a period of five years~\citep{Vallisneri:2014vxa,Abbott:2019ebz}. To identify the signature of gravitational waves from compact-binary mergers, we use matched filtering to extract the signal-to-noise ratio (SNR) of a potential signal~\citep{Allen:2005fk}. Candidates are identified by looking for peaks in the single-detector SNR time series and correlating these triggers between observing detectors. Each candidate is assessed for consistency between the expected signal morphology and the data~\citep{Allen:2004gu,Nitz:2017lco} and ranked using additional factors such as the observed rate of triggers, multi-detector coherence~\citep{Nitz:2017svb,Davies:2020} and the local data quality~\citep{Mozzon:2020gwa,TheLIGOScientific:2017lwt}. This procedure is the same as that used for the prior 3-OGC catalog~\citep{Nitz:2021uxj} and is implemented using the open-source PyCBC toolkit~\citep{pycbc-github,Usman:2015kfa}. This toolkit has also been used to construct many matched-filter based gravitational-wave searches~\citep{LIGOScientific:2018mvr,Abbott:2020niy,LIGOScientific:2021djp,LIGOScientific:2021tfm}, including those that account for eccentricity~\citep{Nitz:2019spj}, are focused on intermediate-mass BBHs~\citep{Chandra:2021wbw}, or target subsolar-mass mergers~\citep{Nitz:2020bdb,Nitz:2021mzz,Nitz:2021vqh}. 

\subsection{LIGO and Virgo Observing Period}

The public LIGO and Virgo data spans the time period from 2015-2020 and the three completed observing runs O1-O3~\citep{Vallisneri:2014vxa,Abbott:2019ebz}. As in 3-OGC~\citep{Nitz:2021uxj}, we include results from data released outside the nominal observation runs, notably this includes data around GW170608~\citep{Abbott:2017gyy} and GW190814~\citep{Abbott:2020khf}. Table~\ref{table:data} shows the number of days that different network configurations were observing. Our updated catalog includes the recently released second half of the third observing run (O3b). The full O3 observing run contains 152 days of triple-detector observing time and $\sim200$ days when both LIGO observatories were observing. We analyze all data when multiple detectors are observing and also identify sources in data when only LIGO-Livingston or LIGO-Hanford are observing. Due to the decreased range and relatively large population of non-Gaussian transient noise, we do not consider data when only Virgo is observing. In Fig.~\ref{fig:range} we show the sky-averaged fiducial BNS range of each instrument as a function of time. All instruments have made significant gains in sensitivity during O3 as compared to O2, with Virgo reaching 60 Mpc range and LIGO-Livingston periodically exceeding 140 Mpc.

The data has been calibrated by the LVK collaborations and made available through the Gravitational-wave Open Science Center (GWOSC)~\citep{Viets:2017yvy, Acernese:2018bfl, Bhattacharjee:2020yxe, Estevez:2020pvj,Vallisneri:2014vxa,Abbott:2019ebz}. Noise subtraction using auxiliary witness channels is applied to the bulk data release~\citep{Davis:2018yrz, Vajente:2019ycy, Estevez2019, Rolland2019}. Our analysis also makes use of the data quality information compiled by the LVK detector characterization teams~\citep{Davis:2021ecd} to exclude times around hardware signal injections and times affected by adverse instrumental behavior.
\vspace{10pt}
\subsection{Search Space}

\begin{figure}[t]
  \centering
    \includegraphics[width=\columnwidth]{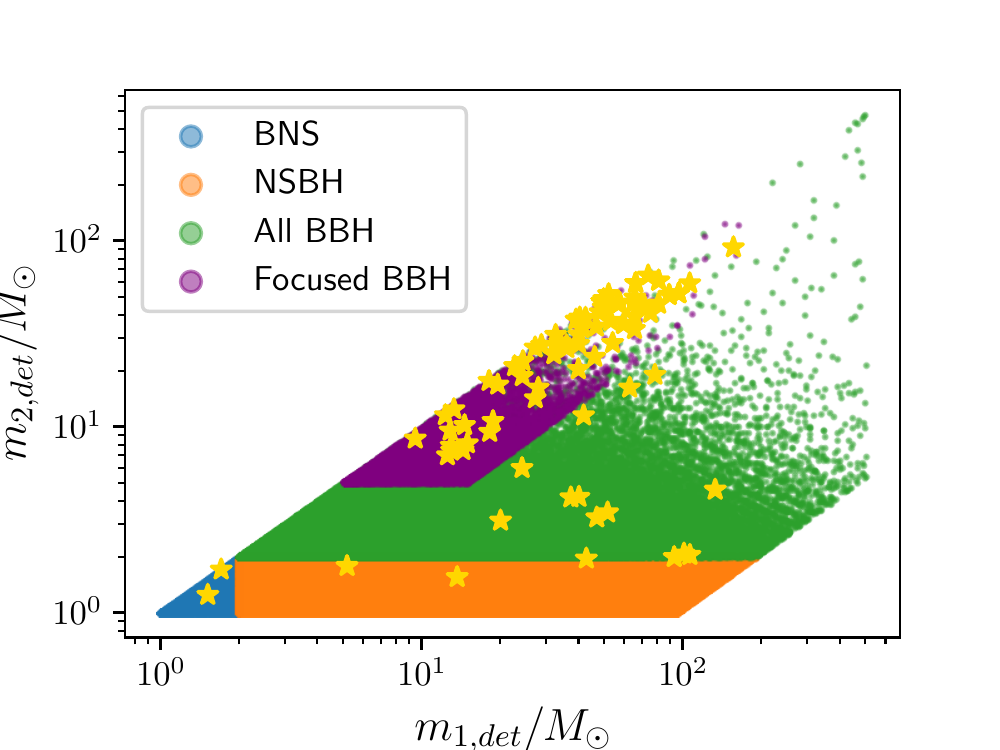}
\caption{The bank of templates used to identify compact-binary mergers in our search as a function of their detector-frame masses. The binary neutron star (blue), neutron star--black hole (orange), binary black hole (green), and focused binary black hole (purple) regions are shown. The templates associated with an observed merger are shown with stars. Signals often have multiple templates which will produce a candidate. Here, we only show the template which produced a candidate with the lowest false alarm rate; the parameters of the selected template can only be considered as crude point estimate of the true parameters, so they may differ significantly from the posterior estimates in Table~\ref{table:peresults}.
}
\label{fig:bank}
\end{figure}

Our search uses matched-filtering to identify signals within the data; matched filtering requires a model of the expected signal to filter the data. To identify sources within a broad target region of parameters (component masses $m_{1,2}$ and spins), we use a discrete bank of template waveforms. For this catalog, we use the same template bank, shown in Fig.~\ref{fig:bank}, that we previously used in the 3-OGC analysis~\citep{Nitz:2021uxj}.

The templates are chosen to ensure that a minimum fraction of the optimal SNR of a potential signal is recovered; typically $\mathcal{O}(97\%)$ for many searches~\citep{DalCanton:2017ala}. The bank is constructed in 4 parts using a brute-force stochastic algorithm~\citep{Harry:2009ea,Ajith:2012mn}. These regions include a focused low-mass-ratio, low-spin, BBH region with a target SNR recovery $>99.5\%$. In addition, there are BNS, NSBH, and broad-parameter BBH banks which target SNR recovery $>97\%$.

The template bank is designed to recover gravitational-wave signals from non-precessing quasi-circular sources. Our search accounts only for the effects of the dominant-mode gravitational-wave signal and does not include the effects of higher-order modes. Neglecting these effects will reduce the search sensitivity to sources which strongly exhibit these features, such as for highly-inclined, high-mass-ratio, or highly precessing binaries~\citep{Harry:2016ijz} or where there remains residual eccentricity~\citep{Ramos-Buades:2020eju, Wang:2021qsu}. Development of optimal search strategies for these sources is an ongoing endeavor~\citep{Harry:2017weg,Harry:2016ijz} and techniques that don't rely on matched filtering also target these sources~\citep{Klimenko:2008fu,Klimenko:2015ypf,Tiwari:2015gal}. To model the gravitational-wave signal, we use a combination of TaylorF2 (for BNS sources)~\citep{Sathyaprakash:1991mt,Droz:1999qx,Blanchet:2002av,Faye:2012we}, SEOBNRv4\_ROM (for BBH and NSBH)~\citep{Taracchini:2013,Bohe:2016gbl}, and IMRPhenomD (focused BBH)~\citep{Husa:2015iqa,Khan:2015jqa}.

\vspace{20pt}
\subsection{Candidate Selection and Significance}
\label{sec:searchsig}
Candidates are assigned a ranking statistic value according to their SNR, consistency with the expected signal morphology~\citep{Allen:2004gu,Nitz:2017lco}, and coherence between observing detectors~\citep{Nitz:2017svb,Davies:2020}. The statistical significance of multi-detector candidates is assessed by comparing to empirically estimated background from artificially produced analyses. By construction these cannot contain astrophysical sources as they are produced by time-shifting the data from one or more detectors by a constant greater than the light-travel-time between detectors~\citep{Babak:2012zx,Usman:2015kfa}. This technique has been used in many past analyses~\citep{Nitz:2019hdf,Venumadhav:2019tad,Colaboration:2011np,Abbott:2009qj,Abbott:2020niy,LIGOScientific:2021djp}. For multi-detector candidates, the estimated background distribution is used to establish the false alarm rate (FAR) of the search as a function of ranking statistic value. Candidates detected in a single detector are instead assessed against off-source observation time. In particular, we use time when both LIGO detectors are observing which allows for confident multi-detector observations to be excised from the background, minimizing potential signal contamination. To limit the effects of non-stationary noise and non-Gaussian noise transients, we further restrict our single-detector analysis to candidates which arise from either the BNS region, focused-BBH with chirp mass $\mathcal{M} < 60$, or NSBH with total mass $M < 50$. The highest mass templates of each region are the most difficult to distinguish from non-Gaussian transient noise due to their short duration.

The probability of astrophysical origin $\mathcal{P}_{\textrm{astro}}$ is calculated using the empirically measured background and comparing to the distribution of observed ranking statistic values produced by a simulated source population as part of a two-component mixture model~\citep{Farr:2015}. For the single-detector analysis, the background distribution is extrapolated using the method of~\citep{Nitz:2020naa}. This is the same methodology as previously used the 3-OGC analysis. We limit our assignment of $\mathcal{P}_{\textrm{astro}}$ to the single-detector analyses using the method of~\citep{Nitz:2020naa} and multi-detector candidates from the focused-BBH region, where the vast majority of observations are found. Due to the uncertain population distribution in other regions, we do not assign a probability of astrophysical origin.

In the 3-OGC analysis, we implicitly assumed a detection-prior that is flat in redshifted chirp-mass~\citep{Nitz:2021uxj} within the focused-BBH region; we make a marginal improvement to this step by averaging the obtained astrophysical probability over this fiducial population scenario with the one obtained from the smoothed observed distribution of redshifted chirp-masses of the high confidence observations (FAR $<$ 1 per 100 yrs). Due to the broad observed distribution in chirp masses, this produces only very modest changes to the estimated astrophysical probability compared to those previously reported in 3-OGC; the primary impact is to mildly increase support for sources with $\mathcal{M} < \sim 50$ in lieu of higher mass sources. A future improvement for the assessment of marginal candidates would be to include a physical model that can simultaneously fit the population along with the foreground and background distributions.
\begin{figure*}[t]
  \centering
    \includegraphics[width=1\columnwidth]{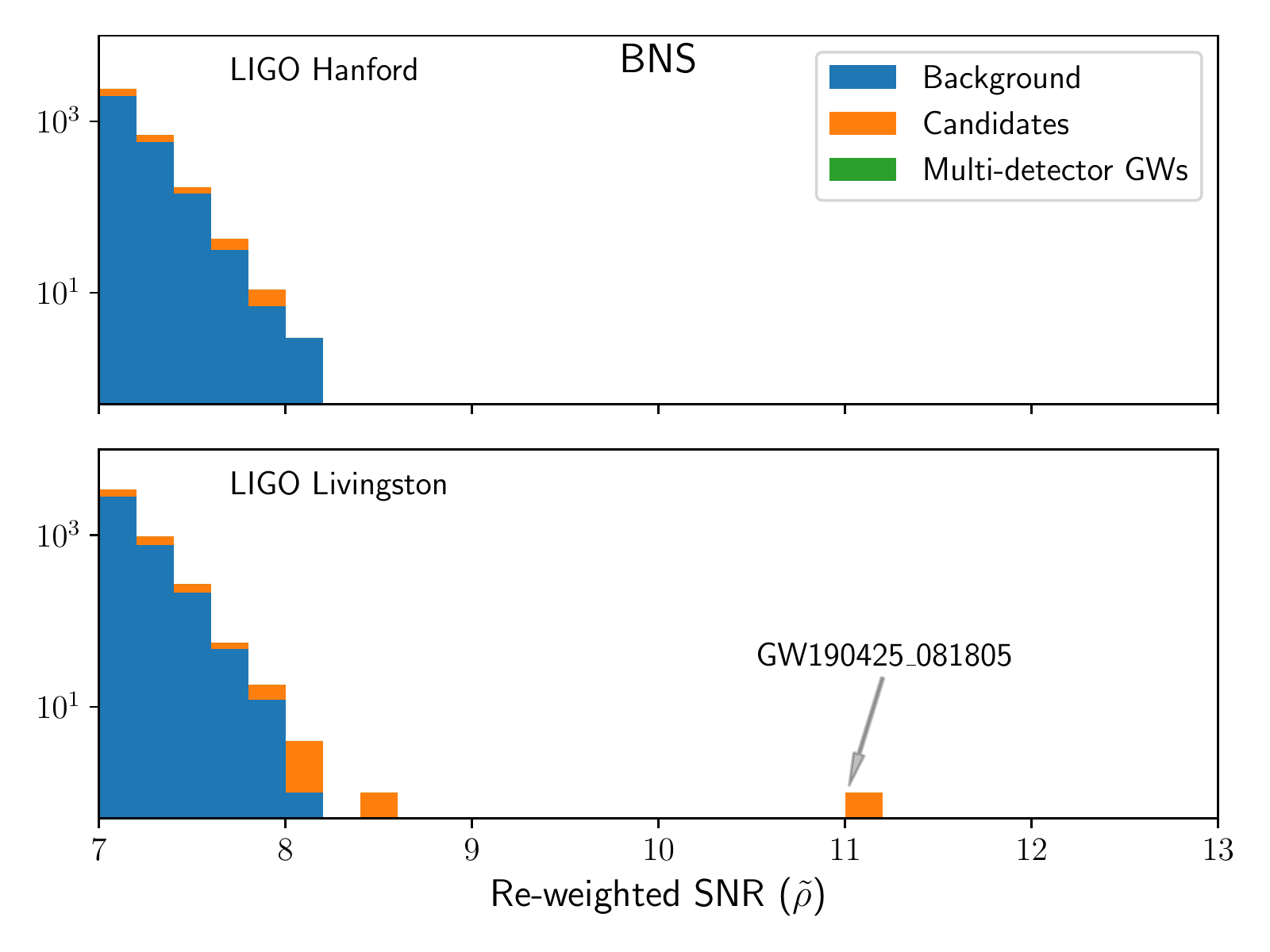}
    \includegraphics[width=1\columnwidth]{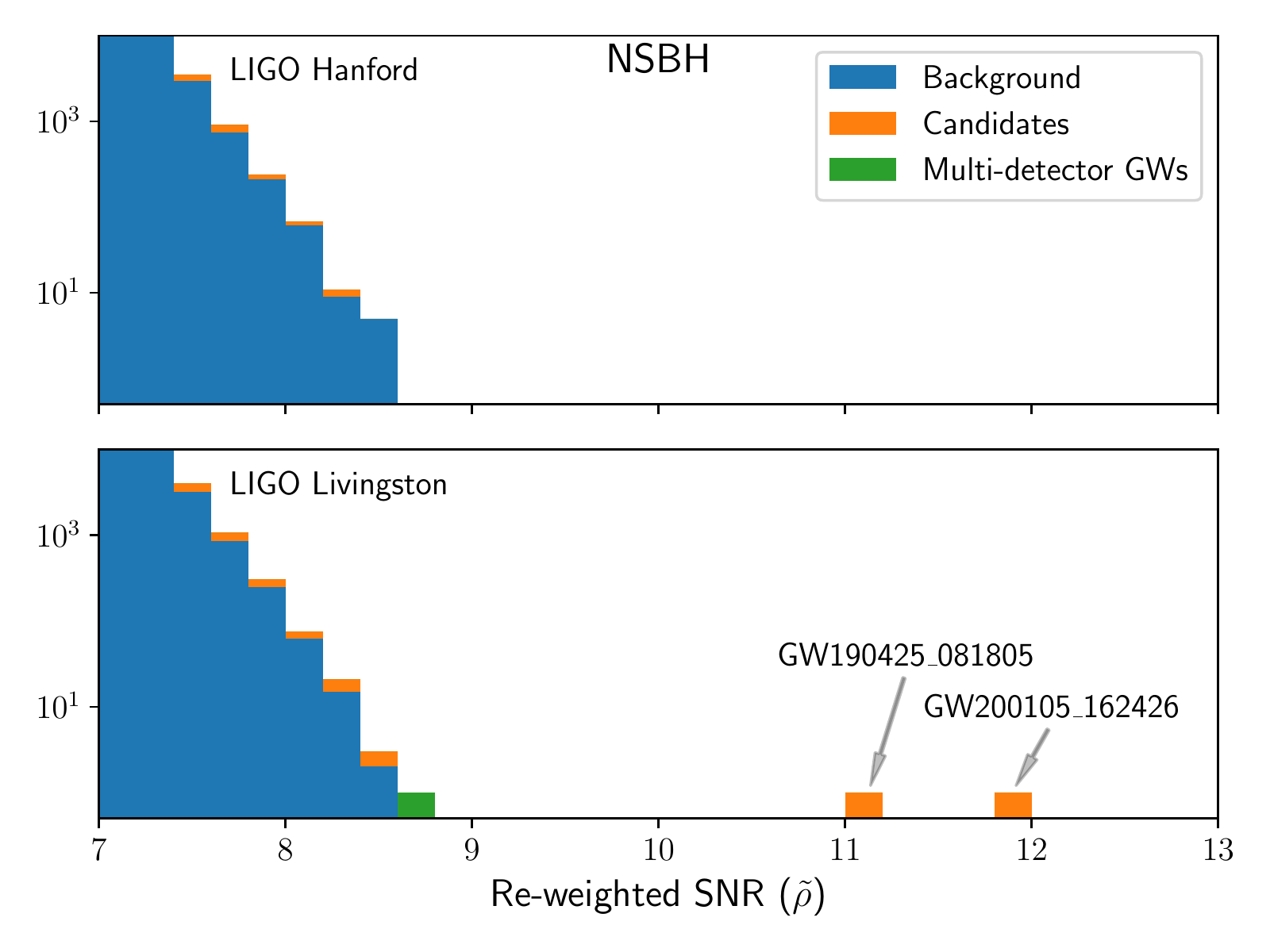}
    \includegraphics[width=1\columnwidth]{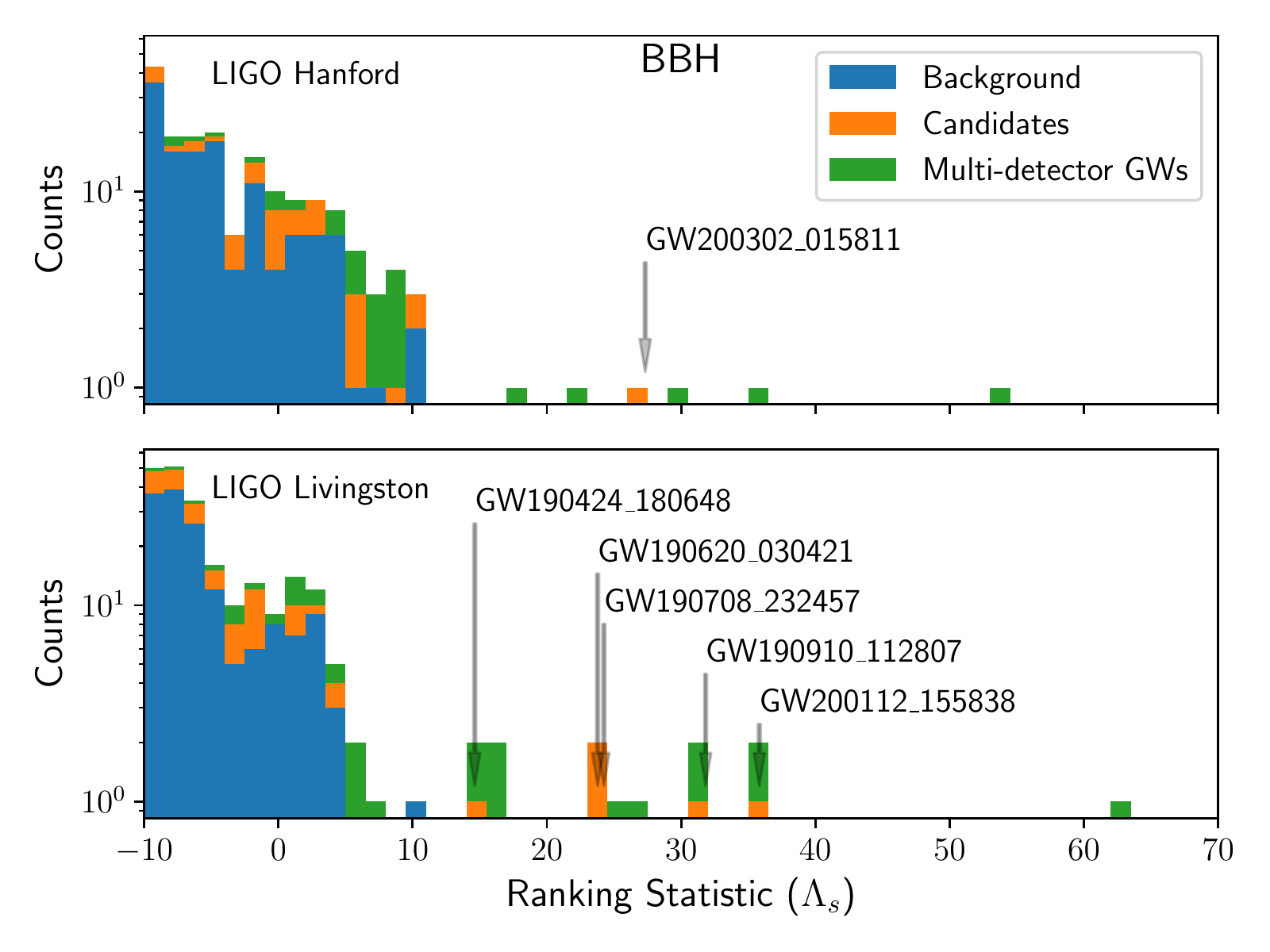}
\caption{The stacked distributions of single-detector triggered candidates observed during O3 when a single LIGO observatory was operating (orange), our selected background (blue), and for comparison the distribution of gravitational-wave mergers observed by the multi-detector analysis (green) as a function of the ranking statistic. Results from the BNS (left), NSBH (right), and BBH (bottom) analyses are shown. The method of~\cite{Nitz:2020naa} is used to extrapolate the background distribution and estimate the probability of astrophysical origin. The BBH analysis uses the statistic $\lambda_s$~\citep{Nitz:2020naa} while the others use a re-weighted SNR statistic~\citep{Babak:2012zx,Nitz:2019hdf}. GW190425 appears in both the BNS and NSBH analysis as candidates were produced from both regions.}
\label{fig:single}
\end{figure*} 

\section{Observational Results}
All events identified by our search with a probability of astrophysical origin $\mathcal{P}_{\textrm{astro}} > 0.5$ or inverse false alarm rate (IFAR) $>$ 100 years are listed in \autoref{table:search}. We report $94$ events, $90$ of which are BBHs, $2$ originate from BNS mergers, and $2$ are the result of coalescing NSBHs.  The table contains information on the observational status of the three different instruments LIGO Hanford, LIGO Livingston, and Virgo, as well as the recovered SNR for the detectors which triggered on the event. Finally, we also provide the Global Positioning System (GPS) times, estimates of the $\mathcal{P}_{\textrm{astro}}$ for BBH signals, and the observed IFARs.

Our search finds 7 new BBH signals during O3b which pass our significance criteria and were not reported as marginal or confident observations by previous studies~\citep{LIGOScientific:2021djp}. These events are GW191224\_043228,  GW200106\_134123, GW200129\_114245, GW200210\_005122, GW200214\_223307, GW200305\_084739, and GW200318\_191337. In addition, 3 previously marginal BBH candidates from the previous 3-OGC catalog~\citep{Nitz:2021uxj} now pass our criteria; this is due to the updated prescription for calculating $\mathcal{P}_{\textrm{astro}}$ (see sec~\ref{sec:searchsig}).  No new BNS or NSBH were discovered at high confidence.

The most confident new detection is GW200318\_191337 with a $\mathcal{P}_{\textrm{astro}} = 0.97$. It is detected in both LIGO Hanford and LIGO Livingston with SNR $\leq 6.2$.  GW200318\_191337 may be the most distant observed merger with redshift  z$\sim0.84^{+0.4}_{-0.35}$. GW200129\_114245 is a potential new mass ratio~$\sim2$ BBH with primary mass $79.1^{+40.2}_{-37.6}\msun$. GW200129\_114245 is detected by both LIGO Hanford and LIGO Livingston with SNR $\leq 6$ and occurs less than $5$ hours after the previous event GW200129\_065458.

The results of our single-detector analysis is shown in Fig.~\ref{fig:single} for BNS, BBH, and now NSBH sources. There have now been 8 mergers detected in a single detector: 6 BBHs, 1 BNS, 1 NSBH. The previously reported NSBH merger GW200105\_162426~\citep{LIGOScientific:2021qlt} is found as a near threshold detection in our analysis. It was observed only in LIGO Livingston with a SNR of $13.3$. LIGO Hanford was not observing at the time. It has $\mathcal{P}_\textrm{astro} \sim 0.5$, similar to GW190425, due to its wide separation from the background distribution. The assumed merger rate for similar NSBH sources is determined by the inclusion of GW200115 whose single-detector LIGO Hanford trigger had a ranking statistic marginally larger than all collected single-detector background.

We present 30 marginal sub-threshold events following the criterion $\mathcal{P}_{\textrm{astro}} > 0.2$ or IFAR $> 0.5$ years in Table \ref{table:sub}.  The majority of these are consistent with BBH mergers except for 170722\_065503 and 191219\_163120 which are consistent with a BNS and a NSBH merger, respectively. In total 12 marginal candidates were previously reported in our most recent 3-OGC catalog; these are 151011\_192749, 170425\_055334, 170704\_202003, 190426\_053949, 190509\_004120, 190530\_030659, 190630\_135302, 190704\_104834, 190707\_071722, 190808\_230535 and 190821\_050019. The remainder of our sub-threshold candidates are included in the corresponding data release~\citep{4-OGC}.

We visually inspect spectrograms of the data around each marginal event to check for non-stationary behavior or an excess of non-Gaussian transient noise which could adversely affect the identified candidate. We observed excess power consistent with a blip glitch~\citep{Cabero:2019orq} in several cases at off-source times. In a few other events we observe slight deviations in power from the expected stationary Gaussian noise in the frequency range 60-110 Hz, e.g. in the Hanford data a few seconds before the time of events (e.g. $200102\_095606$ or $200310\_090144$). Further investigation would be required to determine if potential candidates could be adversely affected in these cases.

Our results are otherwise broadly consistent with those of the recently published GWTC-3 catalog~\citep{LIGOScientific:2021djp}. There are eight events reported in GWTC-3 which do not pass our candidate threshold. These are GW191103\_012549, GW191113\_071753, GW191219\_163120, GW200208\_222617, GW200210\_092254, GW200220\_061928, GW200220\_124850, and GW200322\_091133. Both potential new NSBH candidates GW191219\_163120 and GW200210\_092254 from GWTC-3 are not assigned high confidence by our search. GW191219\_163120 and GW200220\_124850 are detected as near threshold triggers with a $\mathcal{P}_{\textrm{astro}} > 0.2$ or IFAR $>$ 0.5 years. All other events are assigned a lower significance. We expect these small differences in the population of near-threshold events are consistent with differing analysis choices. We note that our candidate threshold is marginally more conservative; GWTC-3 includes candidates with $\mathcal{P}_{\textrm{astro}} > 0.5$ in any of several analyses~\citep{LIGOScientific:2021djp}.

\subsection{Binary Source Parameters}

\begin{figure*}[hbtp]
  \centering
   \includegraphics[scale=0.38]{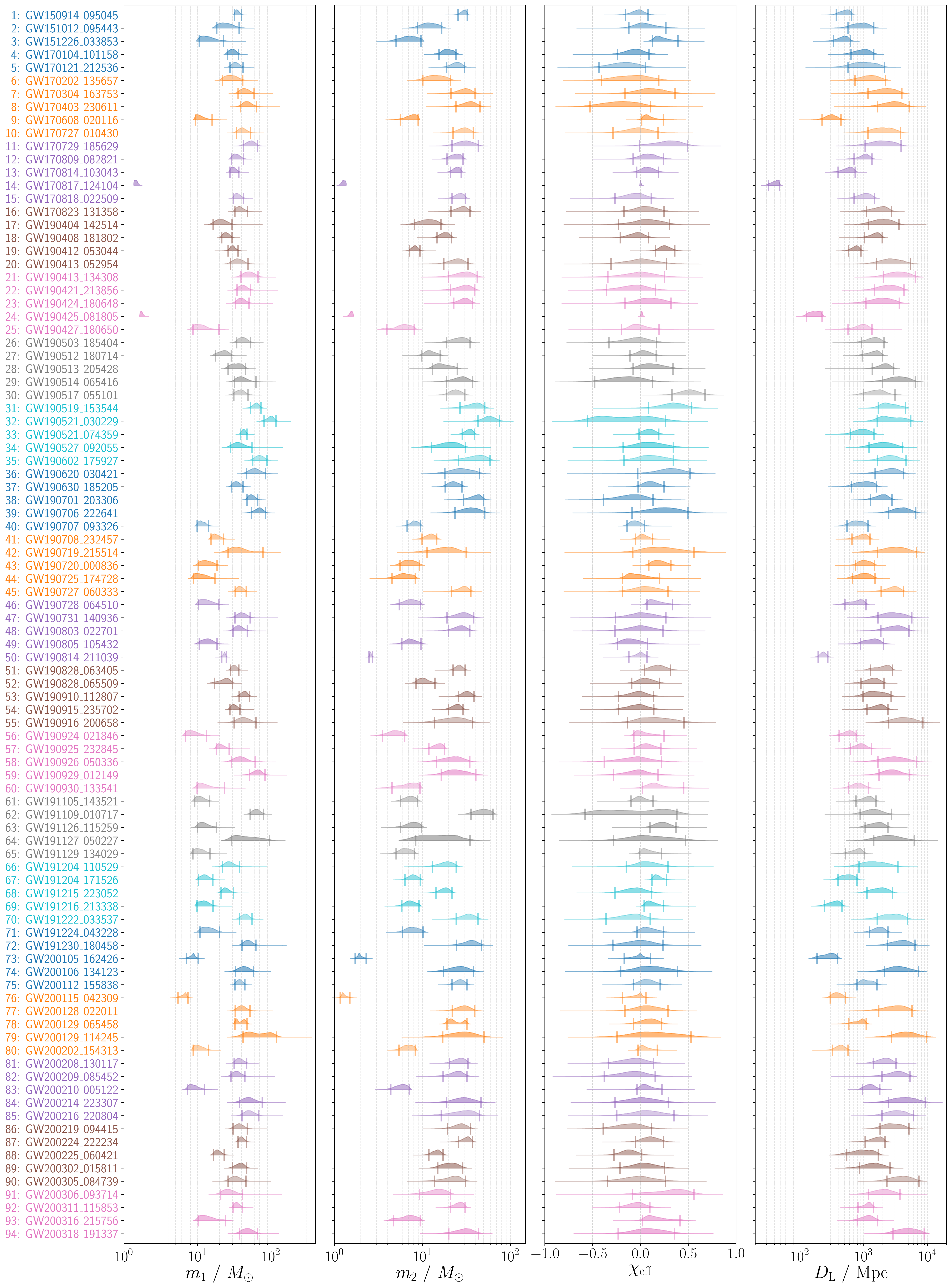}
\caption{The marginalized distributions for component masses $m_1$, $m_2$, the effective spin $\chi_\mathrm{eff}$ and the luminosity distance $D_\mathrm{L}$ for all events that pass our detection criteria in 4-OGC.
The 5th and 95th quantile values are marked with a bar, respectively.
Different colors are used to aid associating each event with its posterior estimates.}
\label{fig:PE-errorbar}
\end{figure*}

In order to infer the properties of the observed compact-binary mergers, we use PyCBC inference \citep{Biwer:2018osg} to perform Bayesian parameter estimation with the standard likelihood assuming the detector noise to be stationary, Gaussian and uncorrelated between the detectors. The parameter estimation results for a total of 94 events are summarized in Fig.~\ref{fig:PE-errorbar} and Table.~\ref{table:peresults}.

\makeatletter\onecolumngrid@push\makeatother
\clearpage
\makeatletter\onecolumngrid@pop\makeatother

For estimating the source parameters of the BBHs, we use IMRPhenomXPHM~\citep{Pratten:2020ceb,lalsuite}, which models precessing binaries and includes higher-order modes. For the two NSBH observations, we use the IMRPhenomNSBH~\citep{Thompson:2020nei} waveform that models a non-precessing NSBH with the neutron star mass, $m_\mathrm{NS}\leq 3 M_{\odot}$, tidal deformability $\Lambda\in[0,5000]$ and the mass ratio no higher than 100~\citep{lalsuite}. For the BNS mergers, we use the IMRPhenomD\_NRTidal waveform model \citep{PhysRevD.93.044007,PhysRevD.93.044006,PhysRevD.96.121501,PhysRevD.99.024029,lalsuite}. This model includes the two tidal deformabilities, $\Lambda_1$ and $\Lambda_2$ of the two components, which characterize the neutron star equation-of-state. 
A dynamical nested sampler \textit{dynesty}~\citep{speagle:2019} is used to sample over the intrinsic parameters mass $m_{1,2}$, spin $\vec{s}_{1,2}$ (in the case of BNS, also tidal deformabilities $\Lambda_{1,2}$, in the case of NSBH, we include the NS tidal deformability $\Lambda_{2}$), and extrinsic parameters luminosity distance $d_{L}$, inclination angle $\iota$, polarization angle $\Psi$, right ascension $\alpha$, declination $\delta$, coalescence time $t_{c}$, and phase $\phi_c$. For BBH sources, the likelihood is marginalized over numerically over $\Psi$, while for BNS and NSBH sources, it is analytically marginalized over $\phi_c$.

We also take into account the effect of calibration uncertainties in amplitude and frequency on the parameter estimation of each event. We use the calibration uncertainty envelopes obtained from the GWOSC for previously known mergers~\citep{Vallisneri:2014vxa}. For the new BBH events, we use the calibration uncertainty envelope from the nearest available time, assuming the calibration envelopes do not change rapidly. The calibration errors in amplitude and phase of the data in each detector are described by frequency-dependent splines~\citep{SplineCalMarg-T1400682} whose parameters are directly sampled over in our analysis. We find that the estimation of intrinsic parameters is not significantly affected by the calibration uncertainties.

The priors for the intrinsic and extrinsic parameters, the \ac{PSD} estimation methods as well as the sampler settings are broadly consistent with those used in the 3-OGC analysis \citep{Nitz:2021uxj}. For the BBH events, uniform priors on source-frame component masses and merger time are used. For the spins, we use uniform priors for the magnitude of the spin and isotropic for the orientation. The distance prior is assumed to be uniform in comoving volume assuming a flat $\Lambda$CDM cosmological model~\citep{Ade:2015xua}. For the rest of the extrinsic parameters, we consider an isotropic prior distribution in the sky localization and binary orientation. 

For the two NSBH systems, the prior on the tidal deformability of the neutron star is considered to be uniform in the range $[0, 5000]$. Other prior distributions are the same as BBH events except that the NS spin magnitude is uniform in $[0, 0.05]$ and the BH is uniform in $[0, 0.5]$. This latter range is chosen based on the parameter space where the waveform model is valid~\citep{Thompson:2020nei, LIGOScientific:2021qlt}.

In the case of BNS mergers, we use the same prior distributions for the component masses, comoving volume, merger time, and orientation, as used in estimating the BBH parameters. However, for GW170817\_124104 we fix the sky location to that from the observed electromagnetic counterpart, $\alpha = 3.446$ and $\delta= -0.408$ \cite{GBM:2017lvd}. The spin magnitudes for the two components are uniform in the range $[0,0.05]$. We also vary the tidal deformability parameters of the NS components, $\Lambda_1$ and $\Lambda_2$ independently in the range $[0, 5000]$. 

\begin{figure*}[t]
  \centering
    \hspace*{-1.8cm}\includegraphics[width=2.6\columnwidth]{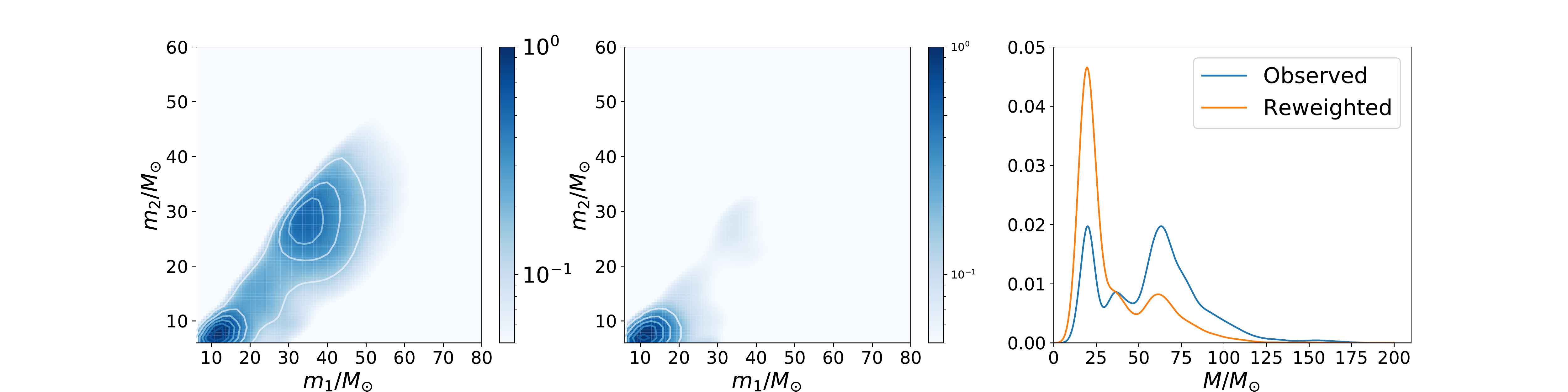}
\caption{Distribution of the source-frame component masses of the BBH population obtained from combining posteriors from all the confident BBH events in the 4-OGC analysis. 
We show the observed component mass distribution (left), component mass distribution corrected for the basic selection effect (middle), and the one-dimensional marginals of the total mass distribution: observed as well as corrected for selection effects (right). To correct for selection effects, we assume a constant detection threshold and correct the distribution for signal loudness varying with component masses.}
\label{fig:bbh_population}
\end{figure*}

\subsection{Binary Black Holes}
A total of 90 BBH mergers have now been observed which pass our significance threshold. 
The growing population will eventually enable distinguishing possible formation channels and the potential contribution from each ~\citep{Talbot:2018cva,LIGOScientific:2018jsj,Roulet:2020wyq,Abbott:2020gyp,LIGOScientific:2021psn}. Notable features expected of the BBH population
were the lower~\citep{Gupta:2019nwj,Zevin:2020gma} and upper~\citep{Fishbach:2020qag,Edelman:2021fik} mass gaps, the former is an observational
expectation from X-ray binaries~\citep{2011ApJ...741..103F,2010ApJ...725.1918O,Bailyn:1997xt}, while the latter is expected due to 
pair-instability supernovae~\citep{Woosley:2016hmi,Marchant:2018kun,Stevenson:2019rcw,vanSon:2020zbk}. 

Several events have components plausibly within the lower mass gap, notably GW190427\_180650, GW190725\_174728, GW190924\_021846, GW190930\_133541, GW200210\_005122 and GW200316\_215756 have posterior support for the secondary mass below 5 $\msun$. GW190814, however, has secondary mass well constrained to be within this region with secondary mass $2.6^{+0.2}_{-0.1}\msun$.
The upper mass gap is expected to occur at $\sim50$-$120$ \msun~\citep{2016MNRAS.457..351Y,Woosley:2016hmi,Belczynski:2016jno,Marchant:2018kun,Woosley_2019,Stevenson:2019rcw,vanSon:2020zbk}. Several detections, including GW190521~\citep{Abbott:2020tfl}, have primary masses within this range. One explanation is that these may be second-generation mergers~\citep{Rodriguez:2019huv,Kimball:2020opk,Gerosa:2021mno}. Other exceptional observations
include the high-mass-ratio GW190814 merger (q~$\sim9$); several proposals exist to explain its formation~\citep{Clesse:2020ghq,Olejak:2020oel,Lu:2020gfh,Carr:2019kxo,Liu:2020gif}.

The observed distribution of source parameters (e.g. masses and spins) and redshift distribution from a large catalog of compact-binary mergers can be used to constrain the population models that predict intrinsic source distribution as well as the rate of mergers~\citep{LIGOScientific:2018jsj,GWTC2-rate,LIGOScientific:2021psn,Roulet:2021hcu,Zhu:2021jbw}. In figure \ref{fig:bbh_population}, we show the distribution of component masses obtained from stacking all the posteriors marginalized over all other parameters, with and without taking into account the zeroth order selection effect that arises due to the loudness of the signal as a function of source parameters. To infer the source population, we use the same method as in the 3-OGC analysis \cite{3-OGC}. We combine the posterior samples from each event to obtain large mass samples. We assume the same priors that we used in the parameter estimation. To account for the loudness of the signal, we estimate the comoving volume corresponding to the ``horizon distance'' of a posterior sample and reassign each sample a weight proportional to the inverse of the comoving volume. The horizon distance is defined as the maximum distance an optimally oriented source can be detected. 

\subsection{BBH Population and Merger Rate}
We use the posterior samples from the individual-source parameter estimation study to infer the distributions of source-frame masses and redshifts for BBHs. The intrinsic distribution of BBH mergers can be parametrized in terms of the `hyper-parameters' $\Theta^{pop}$ of a population model \citep{2020PASA...37...36T, Vitale:2020aaz, GWTC2-rate,LIGOScientific:2021psn}. The hyper parameters describe the shape of the distribution in observed parameters $\theta$ such as masses, spins, redshift, etc. 
To compare the inferred BBH distribution for masses and redshift with other catalogs, we demonstrate using 
models which were also employed in GWTC-3 studies~\citep{LIGOScientific:2021psn}. The sensitivity of our search is estimated by analyzing a simulated population that is injected throughout the O1-O3 data. Our fiducial population includes sources that extend up to a total mass of 300 $M_\odot$ and maximum mass ratio $q$ of 30 with constraints on individual component masses to be between $2M_\odot-150M_\odot$.
In the GWTC-3 analysis, the search sensitivity is estimated semi-analytically for O1+O2 and with direct search performance for O3~\citep{LIGOScientific:2021psn, ligo_scientific_collaboration_and_virgo_2021_5546676, ligo_scientific_collaboration_and_virgo_2021_5636816} using a fiducial BBH population set with component masses between $2M_\odot-100M_\odot$.
We make available the data products needed to perform more detailed follow-up analyses~\citep{4-OGC}; astrophysically-motivated models may provide further insights~\citep{2021ApJ...913L..29F, 2019ApJ...882..121S}. 

\subsubsection{Parametric Models}
The observed mass distribution corrected for first-order selection effects (as shown in figure \ref{fig:bbh_population}) clearly shows a peak in the primary mass between 30-40 $M_\odot$; this peak was previously reported in~\cite{LIGOScientific:2021psn}. 
This observation motivates a parametric model where the primary mass is modelled with a powerlaw component along with added Gaussian components to account for peaks in the mass distribution. 
The powerlaw component of the primary mass is modelled as $p(m_1)\propto m_1^\alpha$. A peak in the observed distribution can be modelled with a Gaussian distribution following \cite{GWTC2-rate,LIGOScientific:2021psn}:
\begin{equation}
    p(m_1) \propto (1-\lambda)\mathcal{C}_{pl}m_1^{-\alpha} + \lambda \mathcal{N}(m_1: \mu^M, \sigma^M), \label{eqn:powerlaw_plus_peak}
\end{equation}
\noindent
where $\lambda$ represents fraction of events in Gaussian component, $\mathcal{C}_{pl}$ is the normalization for power-law component, $\mathcal{N}(x: \mu^M, \sigma^M)$ is the probability density function for a normal distribution for mean $\mu^M$ and standard deviation $\sigma^M$. An additional smoothing function is applied at the low end of the mass distribution in equation \ref{eqn:powerlaw_plus_peak} i.e. $m_{min} <= m_1 < m_{min} + \delta_M$, where $\delta_M$ is the length of the smoothing window. The other population parameters for this model are the minimum $m_{min}$ and maximum $m_{max}$ mass. The conditional probability distribution for the secondary mass $m_2$ is parameterized in terms of mass ratio $q_{pl}=m_2/m_1$ such that $p(q_{pl}|m_1) \propto  q_{pl}^{\beta}$ with same smoothing function at lower end of $m_2$. We distinguish this definition of mass ratio $q_{pl}$ with a subscript `pl' (power-law) from the other definition $q=m_1/m_2$ used elsewhere in the paper. Any additional peak in the primary mass distribution can be added as additional Gaussian component in \ref{eqn:powerlaw_plus_peak}.

The redshift evolution of the BBH merger rate in local comoving time and space coordinates is also given as a power law\citep{2018ApJ...863L..41F}, $R(z) = R_0(1+z)^\kappa$, where $R_0$ defines the local merger rate density at $z=0$ and $\kappa$ represents the power law index. 

\begin{figure*}[t]
  \centering
    \includegraphics[width=2\columnwidth]{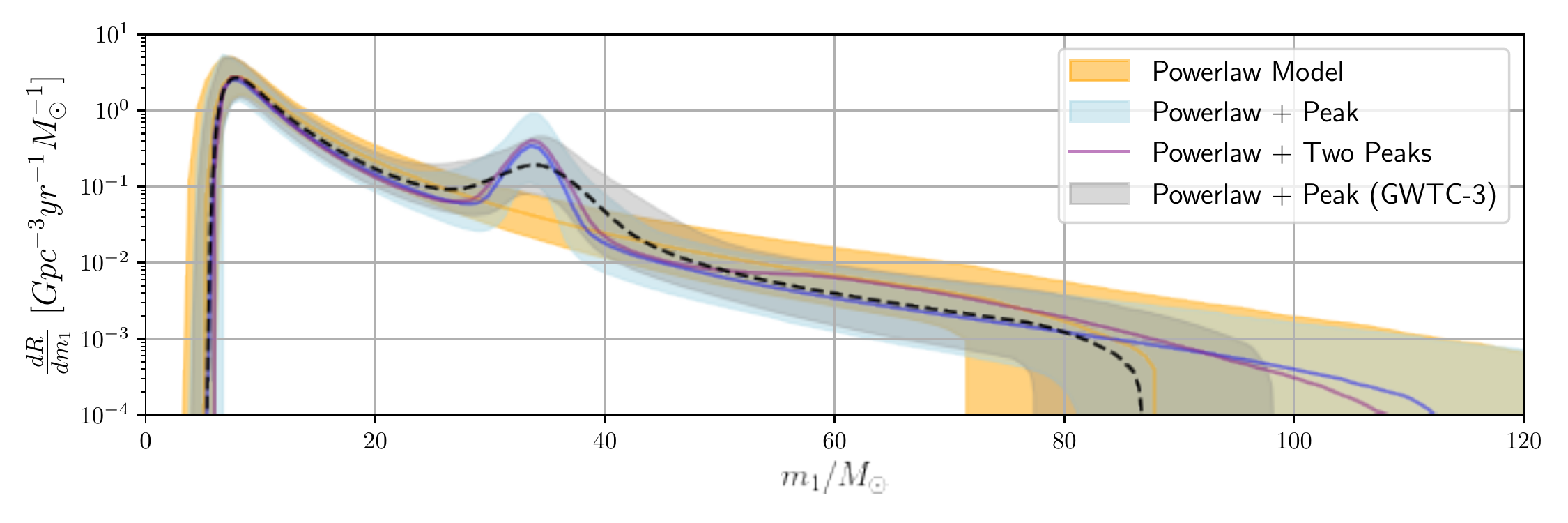}
\caption{Differential merger rate as a function of source-frame primary mass for the parametric mass models considered in the study: power law (orange), powerlaw + peak (blue), and powerlaw + two peaks (purple). We marginalize over all other population parameters in each model. Solid lines show the median value for the analysis done with 4-OGC events. Shaded regions show the 90\% credible intervals except; the powerlaw + two peaks model is omitted to improve visual clarity. For comparison, the powerlaw + peak result from GWTC-3~\citep{LIGOScientific:2021psn} is shown with a black dashed line and grey shaded region.}
\label{fig:differential_merger_rate}
\end{figure*}

\subsubsection{Selection Effects}
A search's sensitivity to BBH mergers depends on various factors such as intrinsic source parameters (such as component masses, spins), extrinsic parameters (such as orientation and sky location of the binary), the detector sensitivity, and behavior of the search to non-Gaussian transient noise.
To infer the intrinsic population distribution, one needs to calculate the detection efficiency of the search i.e. the fraction of the events that can be detected with a population described by the parameters $\Theta^{pop}$. To estimate the detection efficiency, we add a known fiducial population to the data across the observing time of the detector network and use our search to identify them. We use an up-to-date waveform model which includes higher modes and precession, IMRPhenomXPHM~\citep{Pratten:2020ceb,lalsuite}, to simulate the fiducial population.
The reference population follows a power-law mass model on $m_1$ and mass ratio $q_{pl}$ where $p(m_1) \propto m_1^{-\alpha}$ and $p(q_{pl} | m_1) \propto (q_{pl})^\beta$ with $\alpha=2.35, \beta=0$. We use an isotropic spin distribution and the redshift evolution model $p(z)\propto \frac{1}{1+z}\frac{dV_c}{dz}(1+z)^\kappa$ with $\kappa=0$.
The selection function is estimated using a Monte Carlo integral using the sources identified by the search as described in \citep{2018CQGra..35n5009T, 2019RNAAS...3...66F, 2019MNRAS.486.1086M}.

\begin{equation}
\xi(\Theta^{pop}) = \frac{1}{N_{inj}} \sum_{k=1}^{N^{found}_{inj}} \frac{\pi(\theta_k | \Theta^{pop})}{p_{draw}(\theta_k)}, \label{eqn:selection_function}
\end{equation}
\noindent
where $N_{inj}$ are the total number of simulated sources drawn from the fiducial population, $N_{inj}^{found}$ are the number of sources found by the search, $p_{draw}(\theta_k)$ is probability to draw a source from the reference population with parameters $\theta_k$ while $\pi(\theta_k|\Theta^{pop})$ represents the conditional probability distribution of the parameter $\theta_k$ given the population hyper parameter $\Theta^{pop}$.

\begin{figure}[t]
  \centering
    \includegraphics[width=\columnwidth]{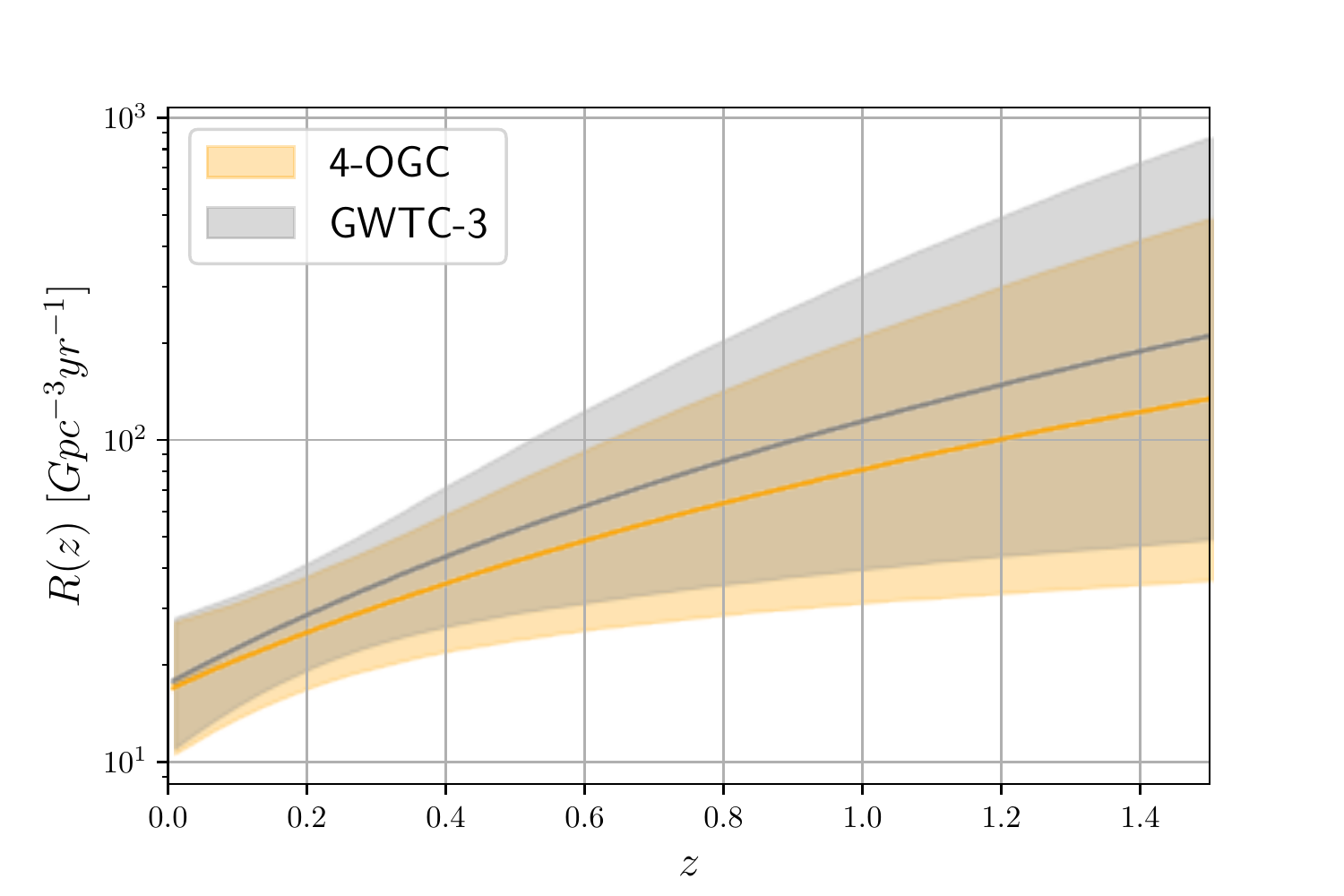}
\caption{The evolution of the merger rate with redshift is shown for powerlaw + peak model along with a comparison to the corresponding GWTC-3 result~\citep{LIGOScientific:2021psn}. The merger rate $R(z)$ is constrained best at around the redshift $z\sim0.2$.}
\label{fig:rate_vs_z}
\end{figure}
\vspace{20pt}

\subsubsection{Parameter Estimation for Population Parameters}
Given the total number of detections $N_{det}$, data ${d}$, population model described by hyper-parameters $\Theta^{pop}$, the population likelihood function can be described by inhomogenous Poisson processes given as \citep{2020PASA...37...36T,2019MNRAS.486.1086M,  Loredo:2004, LIGOScientific:2021psn},
\begin{align}
    \mathcal{L}(\{d\}, N_{det} | & \Theta^{pop}, N_{exp}) \propto\nonumber\\ &N^{N_{det}}e^{-N_{exp}} \prod_{i=1}^{N_{det}}\int\mathcal{L}(d_i|\theta)\pi(\theta|\Theta^{pop})d\theta, \label{pop_likelihood}
\end{align}
\noindent
where $\theta$ is the set of individual source parameters. $\mathcal{L}(d_i|\theta)$ is the likelihood function of $i$-th observation. $N_{exp}$ is the expected number of detections over the full observation period. The total number of mergers $N$ during the same period is given by $N = N_{exp}/\xi(\Theta^{pop})$. 

 We select candidates which pass the following selection criteria: IFAR $>100$ years or $P_{astro}>0.9$. A total of 69 BBH events in 4-OGC catalog satisfy this criteria. We further identify GW190814\_211039 as an outlier from the observed population and exclude it from further analysis due to its very high mass ratio and very low secondary component mass. We use flat priors on all population hyper-parameters except for a log uniform prior on rate parameter $R_0$. We allow the maximum primary mass to vary up to $150 M_\odot$. We use three models for primary mass: i) pure powerlaw component ii) powerlaw + peak , and iii) powerlaw + two peaks. We use a uniform prior for the location of the first peak between $20 M_\odot-45 M_\odot$. For the additional peak in the powerlaw+peak model, we use a uniform prior between $45 M_\odot-80 M_\odot$.

Figure \ref{fig:differential_merger_rate} shows the estimated differential merger rate as a function of the primary mass. The powerlaw+peak model shows a clear peak between $30 M_\odot-40 M_\odot$. The model with two peaks is consistent and does not indicate the presence of a second peak. The results also broadly agree with the GWTC-3 analysis~\citep{LIGOScientific:2021psn}. The small differences can be attributed to the selected observations, differences in the selection function of both searches, and the explored prior ranges. For example, we allow the maximum mass parameter $m_{max}$ to vary up to 150 $M_\odot$ compared to 100 $M_\odot$ in GWTC-3 analysis \citep{LIGOScientific:2021psn}. This explain the differences in the differential merger rate plot at high masses ($ > 100~M_\odot$).

Figure \ref{fig:rate_vs_z} shows the redshift evolution of the BBH merger rate for the powerlaw+peak model. Since the majority of mergers are found near redshift $z\sim0.2$, as expected, the merger rate is best constrained at this redshift. Table \ref{table:rates} shows the merger rates estimated for each model at redshifts $z=0$ and at $z=0.2$.

\begin{table}[]
  \begin{center}
\caption{BBH merger rates (90\% credible interval) for various parametric models at redshift z=0 and 0.2 are shown. We use the parametrization for evolution of merger rates as redshift to be $R(z) = R_0(1+z)^\kappa$. The numbers are quoted for $z=0.2$ along with $z=0$ as the constraints on $R(z)$ are more stronger at z=0.2.}
\label{table:rates}
\begin{tabular}{lcl}
Model & Rate & $[Gpc^{-3} yr^{-1}]$ \\
& z = 0 &  z = 0.2 \\ \hline
  \rule{0pt}{4ex}Power-law    &  $18.47_{-7.00}^{+10.92} $ &  $30.50_{-9.56}^{+15.05}$ \\[0.3cm]  \hline
  \rule{0pt}{4ex}Powerlaw + Peak  &   $16.53^{+10.36}_{-6.19} $  & $24.98^{+12.57}_{-8.06}$ \\[0.3cm]  \hline
  \rule{0pt}{4ex}Powerlaw + Two peaks  &       $17.30^{+10.13}_{-6.65}$     &     $24.36^{+11.89}_{-7.84}$\\[0.3cm] \hline
\end{tabular}
\end{center}
\end{table}

\subsection{Neutron Star Binaries and Neutron Star Black Hole Binaries}
 GW170817\_124104~\citep{TheLIGOScientific:2017qsa} and  GW190425\_081805~\citep{Abbott:2020uma} remain the only confidently observed mergers with BNS-compatible masses. GW170817\_124104 is observed in both the LIGO Hanford and Livingston data with a joint SNR $\sim 33$. This remains the only BNS observation with unambiguous electromagnetic counterparts~\citep{GBM:2017lvd,TheLIGOScientific:2017qsa}. In addition to the $\gamma$-ray burst (GRB 170817A)~\citep{Goldstein:2017mmi,Savchenko:2017ffs,LIGOScientific:2017zic}, the successful electromagnetic follow-up campaign that identified the associated electromagnetic transient near the galaxy NGC 4993~\citep{GBM:2017lvd}, supports the interpretation of this event as a neutron star merger.  GW190425\_081805 is observed only in the LIGO Livingston data with SNR $\sim 11.9$. However, the long duration of the signal increases the power of signal consistency tests~\citep{Usman:2015kfa}; given the distinct separation from the measured background distribution, we consider this a firm detection. It's masses $1.7^{+0.1}_{-0.1}~M_\odot$ and $1.6^{+0.1}_{-0.1}~M_\odot$ (assuming spin magnitude $<$ 0.05) are consistent with interpreting both components as neutron stars. The primary mass is less than the heaviest observed neutron star $\sim 2.2 \msun$~\citep{NANOGrav:2019jur}, however, galactic BNS observations typically have lower mass~\citep{Ozel:2016oaf}. If light black holes in these masses are produced in abundance and form binaries, then we cannot necessarily rule out this explanation from gravitational-wave observation alone. 

We observe two events GW200105\_162426 and GW200115\_042309 with masses consistent with an NSBH system. GW200105\_162426 is observed only in LIGO Livingston data with an SNR $\sim 13.3$. Whereas GW200115\_042309 is observed in both Livingston and Hanford data with IFAR $>$ 100~yr. We are unable to constrain the tidal deformability of the secondary component, which is consistent with previous results~\citep{LIGOScientific:2021qlt,Zhu:2021ysz}.

\subsubsection{Merger Rate}

For neutron star binaries, we use a simplified approach to merger rate estimation. We assume that mergers are Poisson distributed and use a Jeffrey's prior for the rate parameter. We make the simplifying assumption that the observed sources are representative of the mass distribution and do not attempt to fit a mass distribution to the two BNS and two NSBH observations. The observed volume-time for each source class is estimated by simulating the recovery of sources with the same observed source-frame masses and observed low effective spins.

We find that if we consider both GW170817 and GW190425 to be representative members of a common BNS population, we infer the merger rate to be $200^{+309}_{-148}$ Gpc$^{-3}$ yr$^{-1}$ using $90\%$ credible intervals. However, given the unexpectedly large mass of GW190425, it is useful to consider the limits from GW170817 alone under the assumption that these sources either arise from different mechanisms or GW190425 is not a BNS merger. In this scenario, we find that the BNS merger rate is instead $139^{+319}_{-118}$ Gpc$^{-3}$ yr$^{-1}$. Assuming both GW200105 and GW200115 are NSBH mergers, we estimate the merger rate to be $19^{+30}_{-14}$ Gpc$^{-3}$ yr$^{-1}$. The merger rate estimation is dominated by the large uncertainties due to the few observed events and the choice of mass distribution. Alternate choices for the assumed mass distribution can lead to different results; a larger population of observed sources is required to disentangle current rate estimates from assumed mass distributions.

\section{Data Release}

The compilation of supplementary materials is available at~\url{https://github.com/gwastro/4-ogc}~\citep{4-OGC}. To aid in follow-up analyses, the data release contains both the gravitational wave events included in this paper and also the detailed set of sub-threshold BNS, NSBH, and BBH candidates. Auxiliary information such as the parameters of an associated waveform template, ranking statistics, false alarm rate, estimated $\mathcal{P}_{\textrm{astro}}$, results of signal consistency tests, etc. are included where possible. The data release also contains the posterior samples from our parameter estimate for each significant candidate. The configuration files needed to reproduce both the overall search and individual source parameter estimates are provided. We also provide the data products needed to reproduce the BBH population inference with alternate models; this includes the sensitivity of our search to a reference population. Finally, we make available the posterior samples of the parametric population models used in this study.

\section{Conclusions}
4-OGC is a comprehensive catalog of gravitational-wave observations from BNS, NSBH, and BBH mergers covering the 2015-2020 time period. The 94 mergers included in the 4-OGC catalog represent the largest collection to date. These include 90 BBHs, 2 BNS, and 2 NSBH mergers. 7 new BBHs during O3b have been reported here that pass our significance threshold. We estimate the merger rate of BNS, NSBH, and BBH sources at z=0 to be $200^{+309}_{-148}$, $19^{+30}_{-14}$, and $16_{-6}^{+10}$ yr$^{-1}$, respectively. The release of our catalog can help enable deeper investigations of gravitational wave candidates, for instance, to look for deviations from general relativity, study formation channels of compact-binary mergers, and correlate results with other archival observations. 
It is expected that future catalogs may contain hundreds of sources after the ongoing upgrades are completed. The next observing run is scheduled to begin operation at the end of 2022~\citep{Abbott:2020qfu}. The importance of gravitational-wave catalogs will only grow as gravitational-wave astronomy matures and focus is able to shift to understanding increasingly detailed population features.

\acknowledgements
 We are grateful to the computing team from AEI Hannover for their significant technical support. This research has made use of data from the Gravitational Wave Open Science Center (https://www.gw-openscience.org), a service of LIGO Laboratory, the LIGO Scientific Collaboration and the Virgo Collaboration. LIGO is funded by the U.S. National Science Foundation. Virgo is funded by the French Centre National de Recherche Scientifique (CNRS), the Italian Istituto Nazionale della Fisica Nucleare (INFN) and the Dutch Nikhef, with contributions by Polish and Hungarian institutes.

 \setlength{\LTcapwidth}{\textwidth}
\begin{longtable*}{cccccccccc}
  \caption{Shown are the gravitational-wave observations from the full search on data from O1-O3 with $\mathcal{P}_{\textrm{astro}} > 0.5$ or IFAR $>$ 100 years sorted by observation time. We list the GPS time for each event alongside information on the observational status of the three observatories LIGO Hanford (H), LIGO Livingston (L), and Virgo (V). We also list the detectors for which our search triggered and the corresponding recovered SNRs ($\rho$). For some events, where multiple detectors are operational, the search does not trigger for all observatories. This is caused by requiring consistency of the triggers between detectors and triggers in individual detectors needing to exceed the SNR threshold. For multi-detector events we provide the IFAR at the associated ranking statistic for the event. The probability of astrophysical origin is estimated for all events detected by the focused BBH search and from the single-detector analyses. Events reported here for the first time are marked by bold font.}
\label{table:search}\\
 & Event & GPS Time & Observing & Triggered & $\mathcal{P}_\textrm{astro}$ & IFAR [yr] & $\rho_H$ &  $\rho_L$  & $\rho_V$  \\\hline
    \endfirsthead
  \caption*{(Continued) Shown are the gravitational-wave observations from the full search on data from O1-O3 with $\mathcal{P}_{\textrm{astro}} > 0.5$ or IFAR $>$ 100 years sorted by observation time. We list the GPS time for each event alongside information on the observational status of the three observatories LIGO Hanford (H), LIGO Livingston (L), and Virgo (V). We also list the detectors for which our search triggered and the corresponding recovered SNRs ($\rho$). For some events, where multiple detectors are operational, the search does not trigger for all observatories. This is caused by requiring consistency of the triggers between detectors and triggers in individual detectors needing to exceed the SNR threshold. For multi-detector events we provide the IFAR at the associated ranking statistic for the event. The probability of astrophysical origin is estimated for all events detected by the focused BBH search and from the single-detector analyses. Events reported here for the first time are marked by bold font.}
      \label{table:search}\\
 & Event & GPS Time & Observing & Triggered & $\mathcal{P}_\textrm{astro}$ & IFAR [yr] & $\rho_H$ &  $\rho_L$  & $\rho_V$  \\\hline
    \endhead 
    \hline
    \endfoot
 1 & GW150914\_095045 & 1126259462.43 & HL & HL & 1.00 & $>100$ & 19.9 & 13.0 & - \\
\rowcolor{gray!20} 2 & GW151012\_095443 & 1128678900.45 & HL & HL & 1.00 & $>100$ & 6.9 & 6.6 & - \\
\rowcolor{white} 3 & GW151226\_033853 & 1135136350.65 & HL & HL & 1.00 & $>100$ & 10.5 & 7.4 & - \\
\rowcolor{gray!20} 4 & GW170104\_101158 & 1167559936.60 & HL & HL & 1.00 & $>100$ & 8.9 & 9.6 & - \\
\rowcolor{white} 5 & GW170121\_212536 & 1169069154.58 & HL & HL & 1.00 & 16 & 5.2 & 8.9 & - \\
\rowcolor{gray!20} 6 & GW170202\_135657 & 1170079035.73 & HL & HL & 0.86 & 0.50 & 5.4 & 6.2 & - \\
\rowcolor{white} 7 & GW170304\_163753 & 1172680691.37 & HL & HL & 0.74 & 0.25 & 4.6 & 7.0 & - \\
\rowcolor{gray!20} 8 & GW170403\_230611 & 1175295989.23 & HL & HL & 0.72 & 0.25 & 5.2 & 5.5 & - \\
\rowcolor{white} 9 & GW170608\_020116 & 1180922494.49 & HL & HL & 1.00 & $>100$ & 12.4 & 9.0 & - \\
\rowcolor{gray!20} 10 & GW170727\_010430 & 1185152688.03 & HL & HL & 1.00 & 71 & 4.7 & 7.5 & - \\
\rowcolor{white} 11 & GW170729\_185629 & 1185389807.32 & HL & HL & 1.00 & 28 & 7.5 & 6.9 & - \\
\rowcolor{gray!20} 12 & GW170809\_082821 & 1186302519.75 & HLV & HL & 1.00 & $>100$ & 6.7 & 10.7 & - \\
\rowcolor{white} 13 & GW170814\_103043 & 1186741861.53 & HLV & HL & 1.00 & $>100$ & 9.2 & 13.7 & - \\
\rowcolor{gray!20} 14 & GW170817\_124104 & 1187008882.45 & HLV & HL & - & $>100$ & 18.3 & 25.5 & - \\
\rowcolor{white} 15 & GW170818\_022509 & 1187058327.08 & HLV & HL & 1.00 & 5.26 & 4.5 & 9.6 & - \\
\rowcolor{gray!20} 16 & GW170823\_131358 & 1187529256.52 & HL & HL & 1.00 & $>100$ & 6.6 & 9.1 & - \\
\rowcolor{white} 17 & GW190404\_142514 & 1238423132.99 & HL & HL & 0.50 & 0.02 & 5.1 & 5.9 & - \\
\rowcolor{gray!20} 18 & GW190408\_181802 & 1238782700.28 & HLV & HL & 1.00 & $>100$ & 9.2 & 10.3 & - \\
\rowcolor{white} 19 & GW190412\_053044 & 1239082262.17 & HLV & HL & 1.00 & $>100$ & 8.2 & 14.9 & - \\
\rowcolor{gray!20} 20 & GW190413\_052954 & 1239168612.50 & HLV & HL & 1.00 & 1.45 & 5.2 & 6.7 & - \\
\rowcolor{white} 21 & GW190413\_134308 & 1239198206.74 & HLV & HL & 1.00 & 6.39 & 5.4 & 7.8 & - \\
\rowcolor{gray!20} 22 & GW190421\_213856 & 1239917954.25 & HL & HL & 1.00 & $>100$ & 7.9 & 6.3 & - \\
\rowcolor{white} 23 & GW190424\_180648 & 1240164426.14 & L & L & 0.53 & - & - & 9.9 & - \\
\rowcolor{gray!20} 24 & GW190425\_081805 & 1240215503.02 & LV & L & 0.50 & - & - & 11.9 & - \\
\rowcolor{white} 25 & GW190427\_180650 & 1240423628.68 & HLV & HL & 0.53 & 0.02 & 5.8 & 6.8 & - \\
\rowcolor{gray!20} 26 & GW190503\_185404 & 1240944862.29 & HLV & HL & 1.00 & $>100$ & 9.1 & 7.6 & - \\
\rowcolor{white} 27 & GW190512\_180714 & 1241719652.42 & HLV & HL & 1.00 & $>100$ & 5.9 & 10.8 & - \\
\rowcolor{gray!20} 28 & GW190513\_205428 & 1241816086.74 & HLV & HLV & 1.00 & $>100$ & 8.8 & 7.7 & 4.0 \\
\rowcolor{white} 29 & GW190514\_065416 & 1241852074.85 & HL & HL & 0.82 & 0.19 & 6.1 & 5.3 & - \\
\rowcolor{gray!20} 30 & GW190517\_055101 & 1242107479.83 & HLV & HL & 1.00 & 66 & 6.8 & 7.9 & - \\
\rowcolor{white} 31 & GW190519\_153544 & 1242315362.38 & HLV & HL & 1.00 & $>100$ & 7.8 & 9.3 & - \\
\rowcolor{gray!20} 32 & GW190521\_030229 & 1242442967.44 & HLV & HL & 1.00 & $>100$ & 8.4 & 12.0 & - \\
\rowcolor{white} 33 & GW190521\_074359 & 1242459857.47 & HL & HL & 1.00 & $>100$ & 12.1 & 21.0 & - \\
\rowcolor{gray!20} 34 & GW190527\_092055 & 1242984073.79 & HL & HL & 0.94 & 0.37 & 5.0 & 7.0 & - \\
\rowcolor{white} 35 & GW190602\_175927 & 1243533585.10 & HLV & HL & 1.00 & $>100$ & 6.2 & 10.8 & - \\
\rowcolor{gray!20} 36 & GW190620\_030421 & 1245035079.31 & LV & L & 0.73 & - & - & 11.2 & - \\
\rowcolor{white} 37 & GW190630\_185205 & 1245955943.18 & LV & LV & 1.00 & 0.18 & - & 14.7 & 4.0 \\
\rowcolor{gray!20} 38 & GW190701\_203306 & 1246048404.58 & HLV & HLV & 1.00 & 0.13 & 6.0 & 8.9 & 5.7 \\
\rowcolor{white} 39 & GW190706\_222641 & 1246487219.33 & HLV & HL & 1.00 & $>100$ & 9.4 & 8.6 & - \\
\rowcolor{gray!20} 40 & GW190707\_093326 & 1246527224.17 & HL & HL & 1.00 & $>100$ & 7.9 & 9.6 & - \\
\rowcolor{white} 41 & GW190708\_232457 & 1246663515.38 & LV & L & 0.73 & - & - & 12.6 & - \\
\rowcolor{gray!20} 42 & GW190719\_215514 & 1247608532.92 & HL & HL & 0.92 & 0.25 & 5.6 & 5.7 & - \\
\rowcolor{white} 43 & GW190720\_000836 & 1247616534.71 & HLV & HL & 1.00 & $>100$ & 6.8 & 7.7 & - \\
\rowcolor{gray!20} 44 & GW190725\_174728 & 1248112066.46 & HLV & HL & 0.96 & 0.41 & 5.4 & 7.3 & - \\
\rowcolor{white} 45 & GW190727\_060333 & 1248242631.98 & HLV & HL & 1.00 & $>100$ & 7.9 & 8.1 & - \\
\rowcolor{gray!20} 46 & GW190728\_064510 & 1248331528.53 & HLV & HL & 1.00 & $>100$ & 7.5 & 10.6 & - \\
\rowcolor{white} 47 & GW190731\_140936 & 1248617394.64 & HL & HL & 0.92 & 0.43 & 5.2 & 6.0 & - \\
\rowcolor{gray!20} 48 & GW190803\_022701 & 1248834439.88 & HLV & HL & 1.00 & 2.40 & 5.6 & 6.7 & - \\
\rowcolor{white} 49 & GW190805\_105432 & 1249037690.78 & HL & HL & 0.51 & 0.02 & 4.8 & 6.5 & - \\
\rowcolor{gray!20} 50 & GW190814\_211039 & 1249852257.01 & HLV & HL & - & $>100$ & 11.0 & 21.1 & - \\
\rowcolor{white} 51 & GW190828\_063405 & 1251009263.76 & HLV & HL & 1.00 & $>100$ & 10.3 & 11.2 & - \\
\rowcolor{gray!20} 52 & GW190828\_065509 & 1251010527.89 & HLV & HL & 1.00 & $>100$ & 7.3 & 7.4 & - \\
\rowcolor{white} 53 & GW190910\_112807 & 1252150105.32 & LV & L & 0.77 & - & - & 13.4 & - \\
\rowcolor{gray!20} 54 & GW190915\_235702 & 1252627040.70 & HLV & HL & 1.00 & $>100$ & 9.0 & 8.6 & - \\
\rowcolor{white} 55 & GW190916\_200658 & 1252699636.90 & HLV & HL & 0.90 & 0.22 & 4.9 & 5.9 & - \\
\rowcolor{gray!20} 56 & GW190924\_021846 & 1253326744.84 & HLV & HL & 1.00 & $>100$ & 6.7 & 10.8 & - \\
\rowcolor{white} 57 & GW190925\_232845 & 1253489343.12 & HV & HV & 1.00 & $>100$ & 8.2 & - & 5.4 \\
\rowcolor{gray!20} 58 & GW190926\_050336 & 1253509434.07 & HLV & HL & 0.92 & 0.27 & 5.4 & 5.6 & - \\
\rowcolor{white} 59 & GW190929\_012149 & 1253755327.50 & HLV & HL & 0.99 & 3.08 & 5.8 & 7.4 & - \\
\rowcolor{gray!20} 60 & GW190930\_133541 & 1253885759.24 & HL & HL & 1.00 & $>100$ & 6.7 & 7.4 & - \\
\rowcolor{white} 61 & GW191105\_143521 & 1256999739.93 & HLV & HL & 1.00 & $>100$ & 6.0 & 7.7 & - \\
\rowcolor{gray!20} 62 & GW191109\_010717 & 1257296855.21 & HL & HL & 1.00 & $>100$ & 9.5 & 13.0 & - \\
\rowcolor{white} 63 & GW191126\_115259 & 1258804397.63 & HL & HL & 1.00 & 4.86 & 5.8 & 6.7 & - \\
\rowcolor{gray!20} 64 & GW191127\_050227 & 1258866165.55 & HLV & HL & 0.99 & 0.15 & 5.4 & 6.4 & - \\
\rowcolor{white} 65 & GW191129\_134029 & 1259070047.18 & HL & HL & 1.00 & $>100$ & 6.9 & 8.0 & - \\
\rowcolor{gray!20} 66 & GW191204\_110529 & 1259492747.54 & HL & HL & 0.99 & 1.59 & 5.0 & 7.7 & - \\
\rowcolor{white} 67 & GW191204\_171526 & 1259514944.09 & HL & HL & 1.00 & $>100$ & 9.2 & 13.3 & - \\
\rowcolor{gray!20} 68 & GW191215\_223052 & 1260484270.33 & HLV & HL & 1.00 & $>100$ & 7.4 & 7.8 & - \\
\rowcolor{white} 69 & GW191216\_213338 & 1260567236.48 & HV & HV & 1.00 & 55 & 17.5 & - & 5.5 \\
\rowcolor{gray!20} 70 & GW191222\_033537 & 1261020955.12 & HL & HL & 1.00 & $>100$ & 8.6 & 8.2 & - \\
\rowcolor{white} 71 & \textbf{GW191224\_043228} & 1261197166.15 & HLV & HL & 0.87 & 0.13 & 5.0 & 6.8 & - \\
\rowcolor{gray!20} 72 & GW191230\_180458 & 1261764316.40 & HLV & HL & 1.00 & $>100$ & 7.5 & 6.7 & - \\
\rowcolor{white} 73 & GW200105\_162426 & 1262276684.06 & LV & L & 0.50 & - & - & 13.3 & - \\
\rowcolor{gray!20} 74 & \textbf{GW200106\_134123} & 1262353301.93 & HLV & HL & 0.69 & 0.06 & 5.2 & 5.2 & - \\
\rowcolor{white} 75 & GW200112\_155838 & 1262879936.09 & LV & L & 0.78 & - & - & 18.6 & - \\
\rowcolor{gray!20} 76 & GW200115\_042309 & 1263097407.74 & HLV & HL & - & $>100$ & 6.5 & 8.5 & - \\
\rowcolor{white} 77 & GW200128\_022011 & 1264213229.90 & HL & HL & 1.00 & $>100$ & 7.0 & 7.0 & - \\
\rowcolor{gray!20} 78 & GW200129\_065458 & 1264316116.42 & HLV & HV & 1.00 & $>100$ & 14.6 & - & 7.0 \\
\rowcolor{white} 79 & \textbf{GW200129\_114245} & 1264333383.11 & HLV & HL & 0.53 & 0.04 & 5.1 & 6.0 & - \\
\rowcolor{gray!20} 80 & GW200202\_154313 & 1264693411.56 & HLV & HL & 1.00 & 6.05 & 5.1 & 9.6 & - \\
\rowcolor{white} 81 & GW200208\_130117 & 1265202095.95 & HLV & HLV & 1.00 & $>100$ & 6.8 & 6.9 & 4.5 \\
\rowcolor{gray!20} 82 & GW200209\_085452 & 1265273710.17 & HLV & HL & 0.99 & 1.10 & 7.1 & 5.9 & - \\
\rowcolor{white} 83 & \textbf{GW200210\_005122} & 1265331100.74 & HLV & HL & 0.74 & 0.04 & 5.4 & 6.3 & - \\
\rowcolor{gray!20} 84 & \textbf{GW200214\_223307} & 1265754805.00 & HLV & HL & 0.72 & 0.08 & 5.2 & 5.2 & - \\
\rowcolor{white} 85 & GW200216\_220804 & 1265926102.89 & HLV & HL & 0.78 & 0.09 & 6.6 & 5.6 & - \\
\rowcolor{gray!20} 86 & GW200219\_094415 & 1266140673.20 & HLV & HL & 1.00 & 22 & 5.7 & 8.0 & - \\
\rowcolor{white} 87 & GW200224\_222234 & 1266618172.40 & HLV & HL & 1.00 & $>100$ & 12.6 & 13.0 & - \\
\rowcolor{gray!20} 88 & GW200225\_060421 & 1266645879.40 & HL & HL & 1.00 & $>100$ & 9.6 & 7.8 & - \\
\rowcolor{white} 89 & GW200302\_015811 & 1267149509.52 & HV & H & 0.66 & - & 10.5 & - & - \\
\rowcolor{gray!20} 90 & \textbf{GW200305\_084739} & 1267433277.08 & HLV & HL & 0.59 & 0.02 & 4.5 & 6.1 & - \\
\rowcolor{white} 91 & GW200306\_093714 & 1267522652.12 & HL & HL & 0.51 & 0.02 & 5.5 & 5.9 & - \\
\rowcolor{gray!20} 92 & GW200311\_115853 & 1267963151.39 & HLV & HLV & 1.00 & $>100$ & 12.0 & 9.9 & 6.7 \\
\rowcolor{white} 93 & GW200316\_215756 & 1268431094.16 & HLV & HL & 1.00 & 22 & 5.4 & 7.8 & - \\
\rowcolor{gray!20} 94 & \textbf{GW200318\_191337} & 1268594035.14 & HLV & HL & 0.97 & 0.50 & 4.8 & 6.2 & - \\
     %\hline
\end{longtable*}

\begin{longtable*}{ccccccccccccc}
  \caption{The selection of sub-threshold candidates with $\mathcal{P}_{\textrm{astro}} > 0.2$ or IFAR $ > 0.5$ from the full search of O1-O3 data. Candidates are sorted by the observation time. The complete set of sub-threshold candidate is available in the data release and includes a selection of full parameter estimates. Here we show the detector-frame (redshifted) parameters of the template which triggered on the candidate, along with the reported SNRs ($\rho$) from each detector.}
      \label{table:sub}\\
 & Event & GPS Time & Observing & Triggered & $\mathcal{P}_\textrm{astro}$ & IFAR & $\rho_H$ &  $\rho_L$  & $\rho_V$ & $m_{1,\textrm{det}}/\msun$ & $m_{2,\textrm{det}}/\msun$ & $\chi_\mathrm{eff}$\\\hline
 \endfirsthead
   \caption*{(Continued) The selection of sub-threshold candidates with $\mathcal{P}_{\textrm{astro}} > 0.2$ or IFAR $ > 0.5$ from the full search of O1-O3 data. Candidates are sorted by the observation time. The complete set of sub-threshold candidate is available in the data release and includes a selection of full parameter estimates. Here we show the detector-frame (redshifted) parameters of the template which triggered on the candidate, along with the reported SNRs ($\rho$) from each detector.}
    \endhead 
    \hline
    \endfoot
 1 & 151011\_192749 & 1128626886.60 & HL & HL & 0.24 & 0.02 & 4.7 & 6.8 & - & 33.5 & 65.6 & 0.1\\
\rowcolor{gray!20} 2 & 170425\_055334 & 1177134832.19 & HL & HL & 0.37 & 0.07 & 5.3 & 5.8 & - & 46.1 & 65.0 & 0.1\\
\rowcolor{white} 3 & 170623\_234223 & 1182296561.37 & HL & HL & 0.23 & 0.02 & 5.5 & 5.8 & - & 18.2 & 14.4 & 0.1\\
\rowcolor{gray!20} 4 & 170704\_202003 & 1183234821.62 & HL & HL & 0.43 & 0.05 & 5.1 & 6.5 & - & 10.0 & 13.2 & -0.0\\
\rowcolor{white} 5 & 170722\_065503 & 1184741721.32 & HL & HL & - & 0.89 & 5.0 & 7.3 & - & 1.7 & 1.3 & -0.0\\
\rowcolor{gray!20} 6 & 190424\_081138 & 1240128716.76 & HLV & HL & 0.25 & 0.01 & 5.1 & 5.5 & - & 28.4 & 35.1 & 0.0\\
\rowcolor{white} 7 & 190426\_053949 & 1240292407.21 & HLV & HL & 0.40 & 0.01 & 5.2 & 6.1 & - & 20.7 & 20.0 & 0.2\\
\rowcolor{gray!20} 8 & 190509\_004120 & 1241397698.79 & HLV & HL & 0.37 & 0.01 & 4.7 & 6.2 & - & 30.1 & 28.2 & -0.0\\
\rowcolor{white} 9 & 190519\_231324 & 1242342822.07 & HLV & HL & 0.26 & 0.01 & 5.6 & 5.9 & - & 6.6 & 8.7 & -0.0\\
\rowcolor{gray!20} 10 & 190530\_030659 & 1243220837.97 & HLV & HL & 0.44 & 0.01 & 5.2 & 5.8 & - & 26.3 & 45.4 & 0.2\\
\rowcolor{white} 11 & 190630\_135302 & 1245938000.49 & HL & HL & 0.26 & 0.01 & 5.1 & 5.8 & - & 32.6 & 19.2 & 0.0\\
\rowcolor{gray!20} 12 & 190704\_104834 & 1246272532.92 & HLV & HL & 0.27 & 0.01 & 7.0 & 5.5 & - & 5.0 & 5.4 & 0.1\\
\rowcolor{white} 13 & 190707\_071722 & 1246519060.10 & HLV & HL & 0.32 & 0.01 & 6.0 & 5.7 & - & 10.7 & 14.1 & 0.0\\
\rowcolor{gray!20} 14 & 190731\_105943 & 1248606001.71 & HLV & HV & 0.29 & 0.01 & 6.4 & - & 5.2 & 39.1 & 37.6 & -0.0\\
\rowcolor{white} 15 & 190808\_230535 & 1249340753.59 & HLV & HL & 0.41 & 0.01 & 5.0 & 6.5 & - & 13.6 & 13.6 & 0.2\\
\rowcolor{gray!20} 16 & 190821\_050019 & 1250398837.88 & HLV & HL & 0.31 & 0.01 & 5.2 & 5.6 & - & 26.8 & 17.0 & -0.1\\
\rowcolor{white} 17 & 191102\_232120 & 1256772098.02 & HLV & HL & 0.22 & 0.01 & 4.9 & 5.9 & - & 32.0 & 23.1 & -0.0\\
\rowcolor{gray!20} 18 & 191116\_022801 & 1257906499.42 & HLV & HL & 0.22 & 0.00 & 5.2 & 6.4 & - & 12.2 & 7.5 & 0.2\\
\rowcolor{white} 19 & 191122\_214924 & 1258494582.97 & HLV & HL & 0.24 & 0.01 & 4.2 & 7.3 & - & 59.0 & 106.7 & 0.2\\
\rowcolor{gray!20} 20 & 191201\_054144 & 1259214122.47 & HV & HV & - & 0.66 & 7.1 & - & 5.8 & 3.0 & 9.0 & -0.4\\
\rowcolor{white} 21 & 191208\_080334 & 1259827432.84 & HLV & HL & 0.23 & 0.01 & 6.0 & 4.8 & - & 26.7 & 50.1 & -0.1\\
\rowcolor{gray!20} 22 & 191219\_163120 & 1260808298.45 & HLV & HL & - & 0.31 & 5.0 & 7.4 & - & 1.7 & 22.7 & -0.5\\
\rowcolor{white} 23 & 200102\_095606 & 1261994184.05 & HLV & HV & - & 0.47 & 8.9 & - & 4.1 & 123.3 & 77.2 & 0.2\\
\rowcolor{gray!20} 24 & 200116\_082400 & 1263198258.97 & HLV & HL & 0.21 & 0.01 & 4.9 & 5.8 & - & 20.7 & 16.7 & -0.1\\
\rowcolor{white} 25 & 200122\_161511 & 1263744929.34 & HL & HL & 0.27 & 0.01 & 5.7 & 5.7 & - & 6.6 & 13.1 & 0.0\\
\rowcolor{gray!20} 26 & 200205\_141704 & 1264947442.84 & HL & HL & 0.23 & 0.01 & 5.5 & 5.9 & - & 10.3 & 11.1 & 0.2\\
\rowcolor{white} 27 & 200220\_124850 & 1266238148.15 & HL & HL & 0.43 & 0.03 & 6.0 & 5.3 & - & 49.4 & 65.7 & -0.1\\
\rowcolor{gray!20} 28 & 200301\_211019 & 1267132237.66 & HL & HL & 0.43 & 0.02 & 5.5 & 5.9 & - & 31.8 & 17.8 & -0.0\\
\rowcolor{white} 29 & 200304\_182240 & 1267381378.62 & HL & HL & 0.28 & 0.01 & 4.6 & 5.7 & - & 44.3 & 28.4 & -0.1\\
\rowcolor{gray!20} 30 & 200310\_090144 & 1267866122.76 & HL & HL & 0.43 & 0.02 & 5.2 & 6.1 & - & 11.9 & 6.1 & 0.0\\
    %\hline
\end{longtable*}
 
 \setlength{\tabcolsep}{1mm}
\begin{longtable*}[t]{cccccccccccccc}
  \caption{
  We report the Bayesian parameter estimates for all 94 detections from O1-O3 scientific runs and mark the ones reported here for the first time in bold. The various columns report median and 90$\%$ credible intervals for the source-frame component mass $m_1$ and $m_2$, chirp mass $\mathcal{M}$, mass ratio $q$, effective spin $\chi_\mathrm{eff}$, luminosity distance $D_\mathrm{L}$, redshift $z$, and remnant mass and spin $M_f$ and $\chi_f$, respectively. The SNR is computed from the maximum likelihood for BBH events and the maximum likelihood analytically marginalized over the phase for NSBH or BNS events. The quoted results for BBH, BNS and NSBH events are computed using IMRPhenomXPHM, IMRPhenomD\_NRTidal, and IMRPhenomNSBH waveform models, respectively. 
  }\\
  &Event 
& $ {m_1}/M_\odot $ 
& ${m_2}/M_\odot $ 
& $\mathcal{M}/M_\odot $ 
& $q$ 
& $\chi_\mathrm{eff}$ 
& ${D_\mathrm{L}}/\mathrm{Mpc} $ 
& $z$ 
& ${M_\mathrm{f}}/M_\odot$ 
& $\chi_f$ & SNR  \\\hline
  \endfirsthead
   \caption*{(Continued)  We report the Bayesian parameter estimates for all 94 detections from O1-O3 scientific runs and mark the ones reported here for the first time in bold. The various columns report median and 90$\%$ credible intervals for the source-frame component mass $m_1$ and $m_2$, chirp mass $\mathcal{M}$, mass ratio $q$, effective spin $\chi_\mathrm{eff}$, luminosity distance $D_\mathrm{L}$, redshift $z$, and remnant mass and spin $M_f$ and $\chi_f$, respectively. The SNR is computed from the maximum likelihood for BBH events and the maximum likelihood analytically marginalized over the phase for NSBH or BNS events. The quoted results for BBH, BNS and NSBH events are computed using IMRPhenomXPHM, IMRPhenomD\_NRTidal, and IMRPhenomNSBH waveform models, respectively. 
}
      \label{table:peresults}\\
&Event 
& $ {m_1}/M_\odot $ 
& ${m_2}/M_\odot $ 
& $\mathcal{M}/M_\odot $ 
& $q$ 
& $\chi_\mathrm{eff}$ 
& ${D_\mathrm{L}}/\mathrm{Mpc} $ 
& $z$ 
& ${M_\mathrm{f}}/M_\odot$ 
& $\chi_f$ & SNR  \\\hline
    \endhead 
    \hline
    \endfoot
 1& $\mathrm{GW150914\_095045}$ & $34.9^{+4.7}_{-2.9}$ & $29.7^{+2.8}_{-4.2}$ & $27.9^{+1.4}_{-1.3}$ & $1.2^{+0.4}_{-0.2}$ & $-0.03^{+0.1}_{-0.13}$ & $525^{+124}_{-154}$ & $0.11^{+0.02}_{-0.03}$ & $61.6^{+2.9}_{-2.6}$ & $0.69^{+0.05}_{-0.04}$ & 24.0 \\ 
\rowcolor{Gray}2& $\mathrm{GW151012\_095443}$ & $25.2^{+11.8}_{-6.9}$ & $12.5^{+4.1}_{-3.7}$ & $15.1^{+1.3}_{-1.0}$ & $2.0^{+2.1}_{-0.9}$ & $0.04^{+0.22}_{-0.16}$ & $976^{+464}_{-406}$ & $0.19^{+0.08}_{-0.08}$ & $36.4^{+8.5}_{-4.2}$ & $0.65^{+0.09}_{-0.09}$ & 10.2 \\ 
3& $\mathrm{GW151226\_033853}$ & $14.2^{+8.4}_{-3.6}$ & $7.4^{+2.3}_{-2.3}$ & $8.8^{+0.2}_{-0.2}$ & $1.9^{+2.5}_{-0.8}$ & $0.21^{+0.18}_{-0.08}$ & $494^{+129}_{-157}$ & $0.1^{+0.02}_{-0.03}$ & $20.6^{+6.3}_{-1.5}$ & $0.75^{+0.1}_{-0.04}$ & 13.5 \\ 
\rowcolor{Gray}4& $\mathrm{GW170104\_101158}$ & $30.2^{+5.9}_{-4.7}$ & $19.7^{+4.5}_{-4.3}$ & $21.1^{+1.8}_{-1.6}$ & $1.5^{+0.7}_{-0.5}$ & $-0.06^{+0.15}_{-0.18}$ & $1032^{+409}_{-404}$ & $0.2^{+0.07}_{-0.07}$ & $48.0^{+3.7}_{-3.3}$ & $0.66^{+0.07}_{-0.09}$ & 13.8 \\ 
5& $\mathrm{GW170121\_212536}$ & $33.4^{+8.2}_{-5.6}$ & $24.8^{+5.4}_{-5.9}$ & $24.9^{+3.4}_{-3.1}$ & $1.3^{+0.7}_{-0.3}$ & $-0.17^{+0.22}_{-0.26}$ & $1204^{+913}_{-632}$ & $0.23^{+0.15}_{-0.11}$ & $56.0^{+7.5}_{-6.7}$ & $0.64^{+0.09}_{-0.11}$ & 11.0 \\ 
\rowcolor{Gray}6& $\mathrm{GW170202\_135657}$ & $29.7^{+11.7}_{-7.7}$ & $15.0^{+6.0}_{-4.9}$ & $17.9^{+2.7}_{-1.9}$ & $2.0^{+2.0}_{-0.9}$ & $-0.09^{+0.28}_{-0.32}$ & $1379^{+811}_{-645}$ & $0.26^{+0.13}_{-0.11}$ & $43.5^{+8.0}_{-5.5}$ & $0.61^{+0.13}_{-0.15}$ & 9.1 \\ 
7& $\mathrm{GW170304\_163753}$ & $44.8^{+14.0}_{-9.1}$ & $30.9^{+8.9}_{-9.8}$ & $32.0^{+6.2}_{-5.2}$ & $1.4^{+1.1}_{-0.4}$ & $0.1^{+0.26}_{-0.27}$ & $2458^{+1496}_{-1295}$ & $0.43^{+0.21}_{-0.21}$ & $72.3^{+14.0}_{-10.6}$ & $0.73^{+0.1}_{-0.12}$ & 9.0 \\ 
\rowcolor{Gray}8& $\mathrm{GW170403\_230611}$ & $48.3^{+15.0}_{-9.5}$ & $35.2^{+10.1}_{-10.9}$ & $35.4^{+7.7}_{-6.0}$ & $1.3^{+1.0}_{-0.3}$ & $-0.2^{+0.3}_{-0.33}$ & $3208^{+1975}_{-1551}$ & $0.54^{+0.26}_{-0.23}$ & $80.1^{+16.8}_{-12.6}$ & $0.62^{+0.12}_{-0.14}$ & 8.0 \\ 
9& $\mathrm{GW170608\_020116}$ & $11.0^{+4.8}_{-1.6}$ & $7.7^{+1.3}_{-2.1}$ & $8.0^{+0.2}_{-0.2}$ & $1.4^{+1.4}_{-0.4}$ & $0.08^{+0.16}_{-0.06}$ & $322^{+122}_{-99}$ & $0.07^{+0.02}_{-0.02}$ & $17.9^{+2.8}_{-0.6}$ & $0.7^{+0.05}_{-0.04}$ & 15.4 \\ 
\rowcolor{Gray}10& $\mathrm{GW170727\_010430}$ & $41.5^{+11.1}_{-7.6}$ & $30.2^{+7.7}_{-8.0}$ & $30.5^{+5.3}_{-4.2}$ & $1.3^{+0.8}_{-0.3}$ & $-0.03^{+0.21}_{-0.26}$ & $2316^{+1433}_{-1141}$ & $0.41^{+0.2}_{-0.18}$ & $68.6^{+11.4}_{-9.1}$ & $0.68^{+0.09}_{-0.1}$ & 9.0 \\ 
11& $\mathrm{GW170729\_185629}$ & $53.8^{+11.6}_{-11.5}$ & $31.9^{+11.3}_{-10.2}$ & $35.3^{+7.1}_{-5.9}$ & $1.7^{+1.1}_{-0.6}$ & $0.28^{+0.21}_{-0.29}$ & $2290^{+1590}_{-1234}$ & $0.41^{+0.23}_{-0.2}$ & $81.0^{+13.9}_{-11.0}$ & $0.79^{+0.08}_{-0.18}$ & 11.0 \\ 
\rowcolor{Gray}12& $\mathrm{GW170809\_082821}$ & $34.2^{+7.9}_{-5.1}$ & $24.4^{+4.6}_{-5.3}$ & $24.9^{+2.0}_{-1.5}$ & $1.4^{+0.8}_{-0.3}$ & $0.08^{+0.16}_{-0.16}$ & $1064^{+281}_{-352}$ & $0.21^{+0.05}_{-0.06}$ & $55.9^{+4.6}_{-3.4}$ & $0.71^{+0.08}_{-0.08}$ & 12.7 \\ 
13& $\mathrm{GW170814\_103043}$ & $31.2^{+5.3}_{-3.4}$ & $24.7^{+3.0}_{-3.8}$ & $24.0^{+1.3}_{-1.0}$ & $1.3^{+0.5}_{-0.2}$ & $0.07^{+0.12}_{-0.11}$ & $595^{+142}_{-197}$ & $0.12^{+0.03}_{-0.04}$ & $53.2^{+2.8}_{-2.4}$ & $0.71^{+0.06}_{-0.05}$ & 17.7 \\ 
\rowcolor{Gray}14& $\mathrm{GW170817\_124104}$ & $1.5^{+0.1}_{-0.1}$ & $1.3^{+0.1}_{-0.1}$ & $1.186^{+0.002}_{-0.002}$ & $1.2^{+0.2}_{-0.1}$ & $0.0^{+0.01}_{-0.01}$ & $42^{+6}_{-9}$ & $0.01^{+0.0}_{-0.0}$ &  - &  - & 32.7 \\ 
15& $\mathrm{GW170818\_022509}$ & $35.3^{+6.9}_{-4.4}$ & $26.8^{+4.2}_{-5.1}$ & $26.7^{+1.9}_{-1.8}$ & $1.3^{+0.6}_{-0.3}$ & $-0.06^{+0.17}_{-0.21}$ & $1094^{+386}_{-385}$ & $0.21^{+0.07}_{-0.07}$ & $59.6^{+4.2}_{-3.7}$ & $0.68^{+0.07}_{-0.08}$ & 12.0 \\ 
\rowcolor{Gray}16& $\mathrm{GW170823\_131358}$ & $38.2^{+9.8}_{-5.9}$ & $28.7^{+6.2}_{-7.5}$ & $28.5^{+4.1}_{-3.1}$ & $1.3^{+0.8}_{-0.3}$ & $0.05^{+0.2}_{-0.22}$ & $1954^{+794}_{-848}$ & $0.36^{+0.12}_{-0.14}$ & $63.7^{+8.7}_{-6.3}$ & $0.71^{+0.08}_{-0.09}$ & 11.7 \\ 
17& $\mathrm{GW190404\_142514}$ & $21.6^{+7.8}_{-5.3}$ & $12.1^{+4.2}_{-3.9}$ & $13.8^{+2.2}_{-1.9}$ & $1.8^{+1.6}_{-0.7}$ & $0.07^{+0.24}_{-0.3}$ & $2225^{+1358}_{-1046}$ & $0.4^{+0.19}_{-0.17}$ & $32.6^{+5.5}_{-4.6}$ & $0.68^{+0.11}_{-0.18}$ & 8.1 \\ 
\rowcolor{Gray}18& $\mathrm{GW190408\_181802}$ & $24.7^{+4.6}_{-3.4}$ & $18.2^{+3.2}_{-3.5}$ & $18.3^{+1.6}_{-1.1}$ & $1.4^{+0.6}_{-0.3}$ & $-0.04^{+0.13}_{-0.17}$ & $1572^{+391}_{-549}$ & $0.3^{+0.06}_{-0.09}$ & $41.1^{+3.3}_{-2.6}$ & $0.67^{+0.05}_{-0.07}$ & 14.4 \\ 
19& $\mathrm{GW190412\_053044}$ & $30.3^{+5.3}_{-4.1}$ & $8.3^{+1.2}_{-1.1}$ & $13.2^{+0.4}_{-0.3}$ & $3.7^{+1.3}_{-0.9}$ & $0.25^{+0.1}_{-0.09}$ & $758^{+148}_{-181}$ & $0.15^{+0.03}_{-0.03}$ & $37.5^{+4.4}_{-3.2}$ & $0.67^{+0.05}_{-0.04}$ & 19.3 \\ 
\rowcolor{Gray}20& $\mathrm{GW190413\_052954}$ & $36.5^{+12.2}_{-8.2}$ & $25.3^{+7.7}_{-7.7}$ & $26.1^{+5.5}_{-4.6}$ & $1.4^{+1.1}_{-0.4}$ & $0.01^{+0.27}_{-0.32}$ & $3104^{+2334}_{-1511}$ & $0.53^{+0.31}_{-0.23}$ & $59.3^{+12.0}_{-10.0}$ & $0.68^{+0.11}_{-0.13}$ & 8.9 \\ 
21& $\mathrm{GW190413\_134308}$ & $51.3^{+15.6}_{-12.3}$ & $31.2^{+10.8}_{-11.4}$ & $34.0^{+7.1}_{-6.1}$ & $1.6^{+1.5}_{-0.6}$ & $-0.01^{+0.26}_{-0.33}$ & $3955^{+2501}_{-1907}$ & $0.64^{+0.32}_{-0.27}$ & $79.1^{+15.0}_{-12.6}$ & $0.68^{+0.11}_{-0.16}$ & 10.3 \\ 
\rowcolor{Gray}22& $\mathrm{GW190421\_213856}$ & $42.0^{+11.0}_{-7.5}$ & $31.1^{+8.4}_{-10.2}$ & $30.9^{+5.9}_{-4.9}$ & $1.3^{+1.0}_{-0.3}$ & $-0.07^{+0.23}_{-0.29}$ & $2734^{+1542}_{-1247}$ & $0.47^{+0.21}_{-0.19}$ & $69.7^{+12.2}_{-9.6}$ & $0.67^{+0.09}_{-0.13}$ & 10.1 \\ 
23& $\mathrm{GW190424\_180648}$ & $40.1^{+10.2}_{-7.1}$ & $30.6^{+6.9}_{-7.8}$ & $30.2^{+5.0}_{-4.3}$ & $1.3^{+0.7}_{-0.3}$ & $0.1^{+0.22}_{-0.26}$ & $2208^{+1419}_{-1125}$ & $0.4^{+0.2}_{-0.18}$ & $67.2^{+10.9}_{-8.8}$ & $0.73^{+0.09}_{-0.1}$ & 10.5 \\ 
\rowcolor{Gray}24& $\mathrm{GW190425\_081805}$ & $1.7^{+0.1}_{-0.1}$ & $1.6^{+0.1}_{-0.1}$ & $1.431^{+0.015}_{-0.013}$ & $1.1^{+0.2}_{-0.1}$ & $0.01^{+0.01}_{-0.01}$ & $177^{+46}_{-50}$ & $0.04^{+0.01}_{-0.01}$ &  - &  - & 12.4 \\ 
25& $\mathrm{GW190427\_180650}$ & $11.4^{+8.3}_{-2.7}$ & $6.4^{+1.8}_{-2.4}$ & $7.3^{+0.4}_{-0.4}$ & $1.8^{+3.2}_{-0.7}$ & $-0.02^{+0.22}_{-0.17}$ & $983^{+374}_{-412}$ & $0.2^{+0.06}_{-0.08}$ & $17.1^{+6.1}_{-1.4}$ & $0.65^{+0.06}_{-0.09}$ & 9.3 \\ 
\rowcolor{Gray}26& $\mathrm{GW190503\_185404}$ & $42.0^{+10.6}_{-7.9}$ & $27.4^{+7.6}_{-8.5}$ & $29.0^{+4.5}_{-3.8}$ & $1.5^{+1.1}_{-0.5}$ & $-0.04^{+0.21}_{-0.29}$ & $1504^{+575}_{-582}$ & $0.28^{+0.09}_{-0.1}$ & $66.4^{+9.2}_{-7.1}$ & $0.66^{+0.09}_{-0.14}$ & 12.4 \\ 
27& $\mathrm{GW190512\_180714}$ & $23.1^{+5.8}_{-5.5}$ & $12.4^{+3.5}_{-2.6}$ & $14.5^{+1.3}_{-0.9}$ & $1.9^{+1.0}_{-0.7}$ & $0.03^{+0.13}_{-0.14}$ & $1522^{+460}_{-568}$ & $0.29^{+0.07}_{-0.1}$ & $34.2^{+4.0}_{-3.3}$ & $0.66^{+0.07}_{-0.07}$ & 12.4 \\ 
\rowcolor{Gray}28& $\mathrm{GW190513\_205428}$ & $35.2^{+9.8}_{-9.1}$ & $17.6^{+7.4}_{-4.6}$ & $21.2^{+3.2}_{-2.0}$ & $2.0^{+1.3}_{-0.9}$ & $0.11^{+0.22}_{-0.19}$ & $2191^{+772}_{-822}$ & $0.39^{+0.11}_{-0.13}$ & $50.9^{+7.4}_{-5.9}$ & $0.69^{+0.12}_{-0.13}$ & 12.5 \\ 
29& $\mathrm{GW190514\_065416}$ & $41.1^{+21.7}_{-9.3}$ & $28.3^{+9.1}_{-9.5}$ & $29.4^{+7.5}_{-5.3}$ & $1.4^{+1.5}_{-0.4}$ & $-0.16^{+0.28}_{-0.32}$ & $3935^{+2513}_{-1943}$ & $0.64^{+0.32}_{-0.28}$ & $67.0^{+19.5}_{-11.6}$ & $0.63^{+0.11}_{-0.15}$ & 8.4 \\ 
\rowcolor{Gray}30& $\mathrm{GW190517\_055101}$ & $39.0^{+10.0}_{-7.9}$ & $24.2^{+6.1}_{-5.4}$ & $26.6^{+2.8}_{-3.4}$ & $1.6^{+0.9}_{-0.5}$ & $0.51^{+0.16}_{-0.18}$ & $1785^{+1326}_{-787}$ & $0.33^{+0.2}_{-0.13}$ & $59.7^{+7.4}_{-7.7}$ & $0.87^{+0.04}_{-0.05}$ & 11.6 \\ 
31& $\mathrm{GW190519\_153544}$ & $63.1^{+10.1}_{-10.6}$ & $40.3^{+11.1}_{-13.5}$ & $43.3^{+6.2}_{-7.6}$ & $1.6^{+1.0}_{-0.4}$ & $0.34^{+0.2}_{-0.25}$ & $2721^{+1823}_{-1028}$ & $0.47^{+0.25}_{-0.16}$ & $97.6^{+11.6}_{-12.5}$ & $0.79^{+0.08}_{-0.11}$ & 13.9 \\ 
\rowcolor{Gray}32& $\mathrm{GW190521\_030229}$ & $99.8^{+16.9}_{-19.2}$ & $58.6^{+17.1}_{-15.8}$ & $65.4^{+9.8}_{-8.7}$ & $1.7^{+0.8}_{-0.6}$ & $-0.15^{+0.41}_{-0.4}$ & $3147^{+2333}_{-1516}$ & $0.53^{+0.31}_{-0.23}$ & $151.3^{+20.2}_{-15.2}$ & $0.61^{+0.15}_{-0.22}$ & 15.5 \\ 
33& $\mathrm{GW190521\_074359}$ & $42.7^{+4.6}_{-4.1}$ & $34.4^{+4.9}_{-5.9}$ & $33.2^{+2.6}_{-2.7}$ & $1.2^{+0.4}_{-0.2}$ & $0.09^{+0.1}_{-0.1}$ & $1007^{+453}_{-386}$ & $0.2^{+0.08}_{-0.07}$ & $73.2^{+5.2}_{-5.0}$ & $0.71^{+0.04}_{-0.04}$ & 24.5 \\ 
\rowcolor{Gray}34& $\mathrm{GW190527\_092055}$ & $37.1^{+18.0}_{-8.9}$ & $21.6^{+9.5}_{-8.9}$ & $24.0^{+6.6}_{-4.0}$ & $1.7^{+2.3}_{-0.6}$ & $0.09^{+0.25}_{-0.27}$ & $2317^{+1780}_{-1107}$ & $0.41^{+0.25}_{-0.18}$ & $56.6^{+16.7}_{-8.6}$ & $0.7^{+0.12}_{-0.18}$ & 8.9 \\ 
35& $\mathrm{GW190602\_175927}$ & $70.8^{+17.6}_{-14.3}$ & $43.7^{+15.5}_{-18.0}$ & $47.3^{+9.2}_{-9.2}$ & $1.6^{+1.6}_{-0.5}$ & $0.1^{+0.24}_{-0.28}$ & $2847^{+1795}_{-1203}$ & $0.49^{+0.24}_{-0.18}$ & $109.0^{+16.9}_{-14.4}$ & $0.71^{+0.11}_{-0.16}$ & 12.4 \\ 
\rowcolor{Gray}36& $\mathrm{GW190620\_030421}$ & $61.8^{+22.8}_{-15.0}$ & $29.7^{+14.9}_{-11.7}$ & $36.6^{+7.8}_{-6.5}$ & $2.1^{+2.3}_{-0.9}$ & $0.3^{+0.22}_{-0.33}$ & $2821^{+1419}_{-1309}$ & $0.49^{+0.2}_{-0.2}$ & $88.3^{+17.7}_{-12.2}$ & $0.8^{+0.08}_{-0.16}$ & 12.3 \\ 
37& $\mathrm{GW190630\_185205}$ & $34.4^{+7.1}_{-5.1}$ & $22.9^{+5.4}_{-4.7}$ & $24.3^{+2.3}_{-1.8}$ & $1.5^{+0.7}_{-0.4}$ & $0.1^{+0.14}_{-0.13}$ & $1108^{+475}_{-465}$ & $0.22^{+0.08}_{-0.08}$ & $55.0^{+4.8}_{-4.0}$ & $0.7^{+0.06}_{-0.07}$ & 15.6 \\ 
\rowcolor{Gray}38& $\mathrm{GW190701\_203306}$ & $54.8^{+11.3}_{-7.6}$ & $41.6^{+8.3}_{-12.1}$ & $41.0^{+5.3}_{-4.9}$ & $1.3^{+0.8}_{-0.3}$ & $-0.09^{+0.22}_{-0.29}$ & $2027^{+750}_{-738}$ & $0.37^{+0.11}_{-0.12}$ & $91.8^{+10.7}_{-8.7}$ & $0.66^{+0.09}_{-0.12}$ & 12.1 \\ 
39& $\mathrm{GW190706\_222641}$ & $69.6^{+14.4}_{-15.1}$ & $36.8^{+14.4}_{-13.3}$ & $42.7^{+9.2}_{-7.1}$ & $1.9^{+1.5}_{-0.7}$ & $0.24^{+0.26}_{-0.33}$ & $4264^{+2278}_{-1791}$ & $0.68^{+0.29}_{-0.25}$ & $100.8^{+16.4}_{-12.3}$ & $0.78^{+0.1}_{-0.18}$ & 13.0 \\ 
\rowcolor{Gray}40& $\mathrm{GW190707\_093326}$ & $11.6^{+2.7}_{-1.6}$ & $8.2^{+1.3}_{-1.5}$ & $8.5^{+0.4}_{-0.4}$ & $1.4^{+0.7}_{-0.3}$ & $-0.06^{+0.1}_{-0.08}$ & $836^{+332}_{-289}$ & $0.17^{+0.06}_{-0.05}$ & $19.1^{+1.3}_{-1.1}$ & $0.66^{+0.04}_{-0.03}$ & 13.1 \\ 
41& $\mathrm{GW190708\_232457}$ & $18.0^{+4.7}_{-2.7}$ & $12.5^{+2.1}_{-2.5}$ & $12.9^{+0.7}_{-0.5}$ & $1.4^{+0.8}_{-0.4}$ & $0.02^{+0.1}_{-0.07}$ & $987^{+267}_{-332}$ & $0.2^{+0.05}_{-0.06}$ & $29.2^{+2.5}_{-1.6}$ & $0.68^{+0.04}_{-0.04}$ & 13.0 \\ 
\rowcolor{Gray}42& $\mathrm{GW190719\_215514}$ & $38.1^{+40.0}_{-11.9}$ & $19.9^{+11.7}_{-8.6}$ & $23.1^{+14.0}_{-4.3}$ & $1.9^{+3.7}_{-0.8}$ & $0.21^{+0.35}_{-0.29}$ & $3673^{+3066}_{-2024}$ & $0.61^{+0.39}_{-0.3}$ & $55.6^{+41.7}_{-11.3}$ & $0.74^{+0.14}_{-0.21}$ & 8.2 \\ 
43& $\mathrm{GW190720\_000836}$ & $13.5^{+4.9}_{-3.1}$ & $7.3^{+2.0}_{-1.8}$ & $8.6^{+0.4}_{-0.5}$ & $1.8^{+1.4}_{-0.7}$ & $0.19^{+0.13}_{-0.1}$ & $1073^{+402}_{-328}$ & $0.21^{+0.07}_{-0.06}$ & $20.0^{+3.3}_{-1.7}$ & $0.73^{+0.05}_{-0.04}$ & 10.8 \\ 
\rowcolor{Gray}44& $\mathrm{GW190725\_174728}$ & $11.5^{+5.8}_{-2.6}$ & $6.4^{+1.9}_{-1.9}$ & $7.4^{+0.5}_{-0.4}$ & $1.8^{+2.0}_{-0.7}$ & $-0.05^{+0.25}_{-0.14}$ & $1064^{+442}_{-397}$ & $0.21^{+0.08}_{-0.07}$ & $17.3^{+4.0}_{-1.5}$ & $0.65^{+0.08}_{-0.06}$ & 10.0 \\ 
45& $\mathrm{GW190727\_060333}$ & $38.2^{+8.0}_{-5.5}$ & $29.5^{+6.2}_{-8.1}$ & $28.8^{+4.3}_{-3.4}$ & $1.3^{+0.8}_{-0.2}$ & $0.05^{+0.23}_{-0.24}$ & $3057^{+1162}_{-1176}$ & $0.52^{+0.16}_{-0.17}$ & $64.1^{+8.7}_{-6.4}$ & $0.71^{+0.09}_{-0.1}$ & 11.7 \\ 
\rowcolor{Gray}46& $\mathrm{GW190728\_064510}$ & $13.5^{+6.1}_{-3.1}$ & $7.6^{+2.0}_{-2.1}$ & $8.7^{+0.5}_{-0.3}$ & $1.8^{+1.8}_{-0.7}$ & $0.15^{+0.18}_{-0.09}$ & $843^{+247}_{-335}$ & $0.17^{+0.04}_{-0.06}$ & $20.2^{+4.2}_{-1.6}$ & $0.71^{+0.04}_{-0.04}$ & 13.8 \\ 
47& $\mathrm{GW190731\_140936}$ & $40.9^{+11.3}_{-8.4}$ & $29.3^{+9.1}_{-10.2}$ & $29.6^{+6.5}_{-5.6}$ & $1.4^{+1.1}_{-0.3}$ & $0.01^{+0.25}_{-0.27}$ & $3323^{+2345}_{-1638}$ & $0.56^{+0.31}_{-0.24}$ & $66.9^{+13.4}_{-11.1}$ & $0.69^{+0.1}_{-0.13}$ & 8.5 \\ 
\rowcolor{Gray}48& $\mathrm{GW190803\_022701}$ & $37.2^{+10.3}_{-6.8}$ & $27.5^{+7.1}_{-7.7}$ & $27.5^{+5.3}_{-3.9}$ & $1.3^{+0.9}_{-0.3}$ & $-0.0^{+0.24}_{-0.26}$ & $3407^{+1719}_{-1551}$ & $0.57^{+0.23}_{-0.23}$ & $61.6^{+11.4}_{-8.0}$ & $0.69^{+0.09}_{-0.11}$ & 8.8 \\ 
49& $\mathrm{GW190805\_105432}$ & $14.0^{+4.5}_{-3.4}$ & $7.5^{+2.1}_{-1.7}$ & $8.8^{+0.7}_{-0.6}$ & $1.9^{+1.2}_{-0.7}$ & $-0.1^{+0.18}_{-0.14}$ & $1446^{+597}_{-558}$ & $0.27^{+0.1}_{-0.1}$ & $20.7^{+3.3}_{-2.1}$ & $0.62^{+0.05}_{-0.05}$ & 8.9 \\ 
\rowcolor{Gray}50& $\mathrm{GW190814\_211039}$ & $23.3^{+1.4}_{-2.0}$ & $2.6^{+0.2}_{-0.1}$ & $6.1^{+0.1}_{-0.1}$ & $9.0^{+1.0}_{-1.2}$ & $-0.01^{+0.07}_{-0.12}$ & $234^{+39}_{-40}$ & $0.05^{+0.01}_{-0.01}$ & $25.6^{+1.4}_{-1.8}$ & $0.28^{+0.03}_{-0.04}$ & 25.3 \\ 
51& $\mathrm{GW190828\_063405}$ & $31.7^{+4.7}_{-3.7}$ & $26.3^{+4.3}_{-4.1}$ & $24.9^{+3.1}_{-1.9}$ & $1.2^{+0.3}_{-0.2}$ & $0.18^{+0.13}_{-0.14}$ & $2179^{+645}_{-903}$ & $0.39^{+0.1}_{-0.14}$ & $54.7^{+6.6}_{-4.1}$ & $0.75^{+0.06}_{-0.05}$ & 16.1 \\ 
\rowcolor{Gray}52& $\mathrm{GW190828\_065509}$ & $24.3^{+5.6}_{-7.1}$ & $10.4^{+3.5}_{-2.0}$ & $13.5^{+1.2}_{-0.9}$ & $2.3^{+1.1}_{-1.1}$ & $0.05^{+0.15}_{-0.14}$ & $1474^{+603}_{-570}$ & $0.28^{+0.1}_{-0.1}$ & $33.5^{+4.3}_{-4.6}$ & $0.64^{+0.08}_{-0.07}$ & 11.4 \\ 
53& $\mathrm{GW190910\_112807}$ & $43.8^{+7.2}_{-6.8}$ & $32.2^{+6.6}_{-6.5}$ & $32.4^{+3.9}_{-3.6}$ & $1.4^{+0.5}_{-0.3}$ & $-0.03^{+0.16}_{-0.2}$ & $1699^{+1008}_{-726}$ & $0.32^{+0.15}_{-0.12}$ & $72.5^{+8.0}_{-7.5}$ & $0.67^{+0.07}_{-0.09}$ & 13.7 \\ 
\rowcolor{Gray}54& $\mathrm{GW190915\_235702}$ & $31.8^{+6.2}_{-4.1}$ & $24.7^{+4.2}_{-5.0}$ & $24.2^{+2.6}_{-2.0}$ & $1.3^{+0.6}_{-0.2}$ & $-0.03^{+0.17}_{-0.2}$ & $1801^{+612}_{-677}$ & $0.33^{+0.09}_{-0.11}$ & $54.0^{+5.4}_{-4.2}$ & $0.68^{+0.07}_{-0.07}$ & 13.5 \\ 
55& $\mathrm{GW190916\_200658}$ & $43.9^{+18.7}_{-12.3}$ & $24.8^{+12.4}_{-11.1}$ & $27.9^{+8.6}_{-6.4}$ & $1.7^{+2.4}_{-0.6}$ & $0.15^{+0.31}_{-0.29}$ & $4833^{+3600}_{-2366}$ & $0.76^{+0.43}_{-0.32}$ & $66.3^{+17.4}_{-13.9}$ & $0.72^{+0.13}_{-0.21}$ & 7.9 \\ 
\rowcolor{Gray}56& $\mathrm{GW190924\_021846}$ & $8.8^{+4.4}_{-1.9}$ & $5.0^{+1.3}_{-1.5}$ & $5.7^{+0.2}_{-0.2}$ & $1.8^{+2.0}_{-0.7}$ & $0.02^{+0.23}_{-0.09}$ & $598^{+168}_{-184}$ & $0.12^{+0.03}_{-0.04}$ & $13.2^{+3.0}_{-0.9}$ & $0.66^{+0.05}_{-0.05}$ & 12.6 \\ 
57& $\mathrm{GW190925\_232845}$ & $20.9^{+6.1}_{-2.9}$ & $15.3^{+2.5}_{-3.5}$ & $15.5^{+0.9}_{-0.9}$ & $1.4^{+0.9}_{-0.3}$ & $0.06^{+0.15}_{-0.13}$ & $928^{+396}_{-306}$ & $0.19^{+0.07}_{-0.06}$ & $34.7^{+3.1}_{-2.2}$ & $0.7^{+0.06}_{-0.05}$ & 9.9 \\ 
\rowcolor{Gray}58& $\mathrm{GW190926\_050336}$ & $40.0^{+20.9}_{-10.9}$ & $23.8^{+10.7}_{-9.4}$ & $26.2^{+9.3}_{-5.7}$ & $1.7^{+1.7}_{-0.6}$ & $-0.05^{+0.27}_{-0.33}$ & $3575^{+3193}_{-1896}$ & $0.59^{+0.41}_{-0.28}$ & $61.2^{+24.2}_{-12.1}$ & $0.65^{+0.13}_{-0.18}$ & 8.8 \\ 
59& $\mathrm{GW190929\_012149}$ & $66.5^{+16.4}_{-15.7}$ & $25.9^{+14.6}_{-9.5}$ & $35.1^{+8.5}_{-6.7}$ & $2.6^{+2.1}_{-1.2}$ & $-0.03^{+0.22}_{-0.25}$ & $3035^{+2251}_{-1299}$ & $0.52^{+0.3}_{-0.19}$ & $90.2^{+15.1}_{-13.6}$ & $0.58^{+0.16}_{-0.2}$ & 10.3 \\ 
\rowcolor{Gray}60& $\mathrm{GW190930\_133541}$ & $12.7^{+10.6}_{-2.8}$ & $7.5^{+1.9}_{-2.9}$ & $8.4^{+0.4}_{-0.4}$ & $1.7^{+3.4}_{-0.6}$ & $0.16^{+0.29}_{-0.14}$ & $847^{+339}_{-269}$ & $0.17^{+0.06}_{-0.05}$ & $19.3^{+7.8}_{-1.4}$ & $0.73^{+0.08}_{-0.06}$ & 10.2 \\ 
61& $\mathrm{GW191105\_143521}$ & $11.0^{+3.6}_{-1.8}$ & $7.3^{+1.5}_{-1.7}$ & $7.8^{+0.5}_{-0.4}$ & $1.5^{+1.1}_{-0.4}$ & $-0.0^{+0.14}_{-0.09}$ & $1196^{+361}_{-437}$ & $0.23^{+0.06}_{-0.08}$ & $17.7^{+2.0}_{-1.2}$ & $0.67^{+0.05}_{-0.04}$ & 10.2 \\ 
\rowcolor{Gray}62& $\mathrm{GW191109\_010717}$ & $64.0^{+14.9}_{-11.7}$ & $50.1^{+13.5}_{-14.4}$ & $48.4^{+10.8}_{-8.5}$ & $1.3^{+0.6}_{-0.2}$ & $-0.08^{+0.46}_{-0.5}$ & $1555^{+1161}_{-670}$ & $0.29^{+0.18}_{-0.12}$ & $107.7^{+22.0}_{-16.9}$ & $0.68^{+0.18}_{-0.23}$ & 16.5 \\ 
63& $\mathrm{GW191126\_115259}$ & $12.4^{+5.7}_{-2.4}$ & $8.0^{+1.8}_{-2.3}$ & $8.6^{+0.8}_{-0.7}$ & $1.5^{+1.6}_{-0.5}$ & $0.23^{+0.14}_{-0.13}$ & $1710^{+688}_{-653}$ & $0.32^{+0.11}_{-0.11}$ & $19.6^{+3.7}_{-1.9}$ & $0.75^{+0.07}_{-0.05}$ & 8.7 \\ 
\rowcolor{Gray}64& $\mathrm{GW191127\_050227}$ & $52.2^{+42.7}_{-23.0}$ & $19.4^{+15.4}_{-11.0}$ & $25.5^{+12.2}_{-7.1}$ & $2.6^{+6.8}_{-1.5}$ & $0.08^{+0.39}_{-0.37}$ & $3015^{+3202}_{-1581}$ & $0.51^{+0.42}_{-0.24}$ & $70.6^{+37.1}_{-22.1}$ & $0.7^{+0.17}_{-0.32}$ & 10.0 \\ 
65& $\mathrm{GW191129\_134029}$ & $10.9^{+3.9}_{-2.2}$ & $6.6^{+1.6}_{-1.5}$ & $7.3^{+0.4}_{-0.2}$ & $1.7^{+1.2}_{-0.6}$ & $0.07^{+0.15}_{-0.08}$ & $813^{+221}_{-297}$ & $0.16^{+0.04}_{-0.06}$ & $16.8^{+2.4}_{-1.1}$ & $0.69^{+0.04}_{-0.04}$ & 13.2 \\ 
\rowcolor{Gray}66& $\mathrm{GW191204\_110529}$ & $27.8^{+9.6}_{-5.9}$ & $19.2^{+5.2}_{-6.1}$ & $20.0^{+3.1}_{-3.3}$ & $1.4^{+1.3}_{-0.4}$ & $0.07^{+0.22}_{-0.22}$ & $1794^{+1641}_{-953}$ & $0.33^{+0.24}_{-0.16}$ & $45.4^{+6.9}_{-7.5}$ & $0.7^{+0.1}_{-0.1}$ & 9.6 \\ 
67& $\mathrm{GW191204\_171526}$ & $12.7^{+3.5}_{-2.4}$ & $7.9^{+1.6}_{-1.5}$ & $8.7^{+0.3}_{-0.3}$ & $1.6^{+0.9}_{-0.5}$ & $0.17^{+0.1}_{-0.05}$ & $582^{+204}_{-187}$ & $0.12^{+0.04}_{-0.04}$ & $19.7^{+2.1}_{-1.2}$ & $0.73^{+0.04}_{-0.04}$ & 17.6 \\ 
\rowcolor{Gray}68& $\mathrm{GW191215\_223052}$ & $24.6^{+6.1}_{-3.7}$ & $18.3^{+3.5}_{-4.0}$ & $18.3^{+1.9}_{-1.6}$ & $1.3^{+0.8}_{-0.3}$ & $-0.06^{+0.18}_{-0.21}$ & $1935^{+881}_{-807}$ & $0.35^{+0.13}_{-0.13}$ & $41.2^{+4.3}_{-3.8}$ & $0.68^{+0.07}_{-0.07}$ & 11.5 \\ 
69& $\mathrm{GW191216\_213338}$ & $12.4^{+3.9}_{-2.6}$ & $7.4^{+1.8}_{-1.6}$ & $8.3^{+0.2}_{-0.2}$ & $1.7^{+1.1}_{-0.6}$ & $0.11^{+0.13}_{-0.07}$ & $362^{+98}_{-119}$ & $0.08^{+0.02}_{-0.02}$ & $19.0^{+2.4}_{-1.0}$ & $0.7^{+0.04}_{-0.04}$ & 18.6 \\ 
\rowcolor{Gray}70& $\mathrm{GW191222\_033537}$ & $45.1^{+9.7}_{-7.9}$ & $33.9^{+9.1}_{-9.5}$ & $33.5^{+6.5}_{-4.9}$ & $1.3^{+0.8}_{-0.3}$ & $-0.09^{+0.19}_{-0.27}$ & $3147^{+1705}_{-1513}$ & $0.53^{+0.23}_{-0.23}$ & $75.2^{+13.5}_{-10.0}$ & $0.66^{+0.08}_{-0.1}$ & 12.0 \\ 
71& $\textbf{GW191224\_043228}$ & $14.2^{+5.6}_{-3.4}$ & $7.9^{+2.3}_{-2.0}$ & $9.2^{+0.7}_{-0.6}$ & $1.8^{+1.5}_{-0.7}$ & $0.08^{+0.16}_{-0.11}$ & $1804^{+596}_{-576}$ & $0.33^{+0.09}_{-0.09}$ & $21.4^{+3.9}_{-2.1}$ & $0.69^{+0.06}_{-0.05}$ & 9.0 \\ 
\rowcolor{Gray}72& $\mathrm{GW191230\_180458}$ & $49.5^{+12.5}_{-9.3}$ & $36.3^{+11.0}_{-11.7}$ & $36.2^{+8.3}_{-5.9}$ & $1.3^{+0.9}_{-0.3}$ & $-0.01^{+0.25}_{-0.28}$ & $4212^{+2139}_{-1838}$ & $0.68^{+0.27}_{-0.26}$ & $81.3^{+17.2}_{-11.6}$ & $0.69^{+0.1}_{-0.11}$ & 10.4 \\ 
73& $\mathrm{GW200105\_162426}$ & $8.7^{+1.5}_{-1.6}$ & $1.9^{+0.4}_{-0.2}$ & $3.4^{+0.1}_{-0.1}$ & $4.5^{+1.4}_{-1.4}$ & $-0.01^{+0.12}_{-0.16}$ & $298^{+93}_{-113}$ & $0.06^{+0.02}_{-0.02}$ & $10.4^{+1.3}_{-1.3}$ & $0.45^{+0.06}_{-0.02}$ & 13.4 \\ 
\rowcolor{Gray}74& $\textbf{GW200106\_134123}$ & $44.0^{+13.9}_{-10.9}$ & $27.4^{+10.9}_{-9.9}$ & $29.7^{+7.5}_{-6.0}$ & $1.6^{+1.4}_{-0.5}$ & $0.1^{+0.29}_{-0.31}$ & $3989^{+3214}_{-1864}$ & $0.65^{+0.4}_{-0.27}$ & $68.4^{+15.3}_{-13.0}$ & $0.71^{+0.12}_{-0.17}$ & 8.6 \\ 
75& $\mathrm{GW200112\_155838}$ & $37.6^{+6.4}_{-5.3}$ & $27.1^{+5.0}_{-5.3}$ & $27.6^{+2.3}_{-2.2}$ & $1.4^{+0.6}_{-0.3}$ & $0.07^{+0.14}_{-0.14}$ & $1142^{+474}_{-362}$ & $0.22^{+0.08}_{-0.06}$ & $61.8^{+4.6}_{-4.6}$ & $0.7^{+0.06}_{-0.06}$ & 18.6 \\ 
\rowcolor{Gray}76& $\mathrm{GW200115\_042309}$ & $6.6^{+0.8}_{-1.2}$ & $1.3^{+0.2}_{-0.1}$ & $2.4^{+0.0}_{-0.1}$ & $5.2^{+1.1}_{-1.6}$ & $-0.03^{+0.08}_{-0.16}$ & $393^{+123}_{-93}$ & $0.08^{+0.02}_{-0.02}$ & $7.8^{+0.7}_{-1.0}$ & $0.4^{+0.07}_{-0.02}$ & 10.9 \\ 
77& $\mathrm{GW200128\_022011}$ & $40.7^{+11.2}_{-8.0}$ & $30.3^{+9.1}_{-8.4}$ & $30.1^{+7.1}_{-4.9}$ & $1.3^{+0.8}_{-0.3}$ & $0.07^{+0.22}_{-0.23}$ & $3609^{+2097}_{-1885}$ & $0.6^{+0.27}_{-0.28}$ & $67.3^{+15.1}_{-10.3}$ & $0.72^{+0.09}_{-0.1}$ & 10.2 \\ 
\rowcolor{Gray}78& $\mathrm{GW200129\_065458}$ & $40.3^{+6.6}_{-7.8}$ & $23.4^{+8.7}_{-4.4}$ & $26.7^{+2.5}_{-2.0}$ & $1.7^{+0.7}_{-0.7}$ & $0.09^{+0.13}_{-0.16}$ & $911^{+189}_{-327}$ & $0.18^{+0.03}_{-0.06}$ & $61.5^{+4.0}_{-3.3}$ & $0.76^{+0.07}_{-0.06}$ & 27.1 \\ 
79& $\textbf{GW200129\_114245}$ & $79.1^{+40.2}_{-37.6}$ & $31.5^{+18.6}_{-14.4}$ & $41.0^{+15.7}_{-13.1}$ & $2.3^{+3.7}_{-1.2}$ & $0.13^{+0.4}_{-0.38}$ & $5294^{+4287}_{-2561}$ & $0.82^{+0.51}_{-0.34}$ & $109.6^{+34.9}_{-42.4}$ & $0.73^{+0.16}_{-0.21}$ & 8.3 \\ 
\rowcolor{Gray}80& $\mathrm{GW200202\_154313}$ & $10.6^{+3.6}_{-1.8}$ & $7.0^{+1.4}_{-1.5}$ & $7.4^{+0.2}_{-0.2}$ & $1.5^{+1.1}_{-0.5}$ & $0.04^{+0.14}_{-0.06}$ & $442^{+130}_{-120}$ & $0.09^{+0.03}_{-0.02}$ & $16.8^{+2.1}_{-0.7}$ & $0.68^{+0.03}_{-0.03}$ & 11.2 \\ 
81& $\mathrm{GW200208\_130117}$ & $37.6^{+9.0}_{-6.0}$ & $27.2^{+5.9}_{-6.8}$ & $27.5^{+3.4}_{-2.9}$ & $1.4^{+0.8}_{-0.3}$ & $-0.07^{+0.21}_{-0.26}$ & $2259^{+952}_{-844}$ & $0.4^{+0.14}_{-0.13}$ & $62.2^{+7.1}_{-6.1}$ & $0.66^{+0.08}_{-0.11}$ & 11.0 \\ 
\rowcolor{Gray}82& $\mathrm{GW200209\_085452}$ & $34.8^{+9.3}_{-6.2}$ & $25.3^{+7.0}_{-8.0}$ & $25.4^{+5.0}_{-3.9}$ & $1.4^{+1.0}_{-0.3}$ & $-0.08^{+0.24}_{-0.3}$ & $3573^{+1779}_{-1631}$ & $0.59^{+0.23}_{-0.24}$ & $57.2^{+10.8}_{-7.4}$ & $0.67^{+0.09}_{-0.13}$ & 9.4 \\ 
83& $\textbf{GW200210\_005122}$ & $8.9^{+3.5}_{-1.6}$ & $5.9^{+1.2}_{-1.5}$ & $6.3^{+0.3}_{-0.4}$ & $1.5^{+1.3}_{-0.5}$ & $0.06^{+0.17}_{-0.09}$ & $1299^{+490}_{-343}$ & $0.25^{+0.08}_{-0.06}$ & $14.3^{+2.1}_{-1.1}$ & $0.69^{+0.06}_{-0.05}$ & 8.5 \\ 
\rowcolor{Gray}84& $\textbf{GW200214\_223307}$ & $51.6^{+24.4}_{-14.2}$ & $30.9^{+15.5}_{-12.0}$ & $34.2^{+12.5}_{-7.9}$ & $1.6^{+1.8}_{-0.5}$ & $0.01^{+0.28}_{-0.28}$ & $5161^{+4122}_{-2714}$ & $0.8^{+0.49}_{-0.37}$ & $79.6^{+28.8}_{-17.8}$ & $0.68^{+0.12}_{-0.17}$ & 7.9 \\ 
85& $\mathrm{GW200216\_220804}$ & $51.7^{+16.5}_{-11.5}$ & $31.6^{+12.9}_{-15.2}$ & $34.2^{+8.1}_{-8.2}$ & $1.6^{+2.3}_{-0.5}$ & $0.06^{+0.29}_{-0.31}$ & $3643^{+2367}_{-1708}$ & $0.6^{+0.3}_{-0.25}$ & $79.5^{+15.7}_{-12.4}$ & $0.7^{+0.12}_{-0.19}$ & 8.7 \\ 
\rowcolor{Gray}86& $\mathrm{GW200219\_094415}$ & $37.7^{+10.1}_{-7.2}$ & $27.6^{+7.6}_{-8.3}$ & $27.7^{+5.3}_{-4.3}$ & $1.3^{+0.9}_{-0.3}$ & $-0.11^{+0.23}_{-0.28}$ & $3214^{+1919}_{-1396}$ & $0.54^{+0.25}_{-0.21}$ & $62.5^{+11.3}_{-8.8}$ & $0.65^{+0.09}_{-0.12}$ & 10.5 \\ 
87& $\mathrm{GW200224\_222234}$ & $40.1^{+6.2}_{-4.6}$ & $32.2^{+4.9}_{-6.6}$ & $30.9^{+3.1}_{-2.4}$ & $1.2^{+0.5}_{-0.2}$ & $0.1^{+0.14}_{-0.15}$ & $1683^{+474}_{-624}$ & $0.31^{+0.07}_{-0.1}$ & $68.3^{+6.2}_{-4.4}$ & $0.73^{+0.06}_{-0.07}$ & 19.1 \\ 
\rowcolor{Gray}88& $\mathrm{GW200225\_060421}$ & $19.1^{+4.0}_{-2.6}$ & $14.7^{+2.5}_{-2.9}$ & $14.5^{+1.4}_{-1.1}$ & $1.3^{+0.6}_{-0.3}$ & $-0.12^{+0.14}_{-0.15}$ & $1103^{+598}_{-560}$ & $0.22^{+0.1}_{-0.1}$ & $32.4^{+3.0}_{-2.7}$ & $0.66^{+0.06}_{-0.06}$ & 12.9 \\ 
89& $\mathrm{GW200302\_015811}$ & $38.5^{+8.2}_{-8.4}$ & $21.5^{+7.2}_{-6.5}$ & $24.5^{+3.9}_{-3.5}$ & $1.8^{+1.1}_{-0.7}$ & $0.03^{+0.22}_{-0.25}$ & $1582^{+996}_{-723}$ & $0.3^{+0.15}_{-0.12}$ & $57.5^{+7.7}_{-7.0}$ & $0.68^{+0.11}_{-0.14}$ & 11.2 \\ 
\rowcolor{Gray}90& $\textbf{GW200305\_084739}$ & $33.8^{+12.1}_{-7.7}$ & $23.2^{+7.7}_{-9.4}$ & $24.0^{+5.9}_{-5.4}$ & $1.4^{+1.5}_{-0.4}$ & $-0.02^{+0.29}_{-0.32}$ & $4422^{+2903}_{-2113}$ & $0.7^{+0.36}_{-0.29}$ & $54.5^{+13.1}_{-10.2}$ & $0.68^{+0.12}_{-0.15}$ & 8.1 \\ 
91& $\mathrm{GW200306\_093714}$ & $27.5^{+13.5}_{-6.9}$ & $15.4^{+5.4}_{-6.1}$ & $17.7^{+2.9}_{-3.1}$ & $1.8^{+2.4}_{-0.7}$ & $0.32^{+0.24}_{-0.41}$ & $2249^{+1478}_{-1056}$ & $0.4^{+0.21}_{-0.17}$ & $41.3^{+9.4}_{-6.5}$ & $0.79^{+0.09}_{-0.22}$ & 8.1 \\ 
\rowcolor{Gray}92& $\mathrm{GW200311\_115853}$ & $34.6^{+6.1}_{-3.9}$ & $26.5^{+4.3}_{-5.5}$ & $26.1^{+2.2}_{-1.7}$ & $1.3^{+0.6}_{-0.3}$ & $-0.04^{+0.14}_{-0.17}$ & $1187^{+245}_{-370}$ & $0.23^{+0.04}_{-0.07}$ & $58.3^{+4.4}_{-3.3}$ & $0.67^{+0.07}_{-0.07}$ & 17.7 \\ 
93& $\mathrm{GW200316\_215756}$ & $13.9^{+10.2}_{-3.7}$ & $7.3^{+2.2}_{-2.6}$ & $8.7^{+0.5}_{-0.5}$ & $1.9^{+3.2}_{-0.8}$ & $0.14^{+0.27}_{-0.12}$ & $1206^{+452}_{-412}$ & $0.23^{+0.08}_{-0.07}$ & $20.4^{+7.8}_{-2.1}$ & $0.71^{+0.07}_{-0.05}$ & 10.6 \\ 
\rowcolor{Gray}94& $\textbf{GW200318\_191337}$ & $49.1^{+16.4}_{-12.0}$ & $31.6^{+12.0}_{-11.6}$ & $33.5^{+8.8}_{-6.4}$ & $1.5^{+1.4}_{-0.5}$ & $0.08^{+0.28}_{-0.31}$ & $5513^{+3372}_{-2617}$ & $0.84^{+0.4}_{-0.35}$ & $76.9^{+18.8}_{-13.7}$ & $0.71^{+0.11}_{-0.17}$ & 8.4 \\ 
     \hline
\end{longtable*}

\pagebreak
\bibliography{references}

\end{CJK*}
\end{document}